\documentclass[]{aa} 
\usepackage{natbib}           
\usepackage{graphicx}         
\usepackage{amssymb}          
\usepackage[usenames,dvipsnames,svgnames]{xcolor}
\usepackage{txfonts}

\newcommand{\rmd}{{\rm d}}
\newcommand{\cf}{\textit{c.f.}}
\newcommand{\eg}{\textit{e.g.}, }
\newcommand{\ie}{\textit{i.e.}, }

\newlength{\imsize}
\newcommand{\Nabla}{\vec{\nabla}}
\newcommand{\pardiff}[2]{\frac{\partial #1}{\partial #2}}

\newcommand{\dS}{\rmd \vec{S}}
\newcommand{\dV}{\, \rmd \mathcal{V}}
\newcommand{\surf}{{\partial \mathcal{V}}}
\renewcommand{\vec}[1]{ {\mathbf #1} }
\newcommand{\vol}{\mathcal{V}}

\newcommand{\eq}[1]{Eq.~(\ref{eq:#1})} 
\newcommand{\eqs}[2]{Eqs.~(\ref{eq:#1},\ref{eq:#2})} 
\newcommand{\eqss}[2]{Eqs.~(\ref{eq:#1} - \ref{eq:#2})}
\newcommand{\sect}[1]{Sect.~\ref{s:#1}}

\newcommand{\fig}[1]{Fig.~\ref{Fig:#1}} 
\newcommand{\figs}[2]{Figs.~\ref{Fig:#1} and~\ref{Fig:#2}}

\newcommand{\BE}{\begin{equation}}
\newcommand{\EE}{\end{equation}}
\newcommand{\BA}{\begin{eqnarray}}
\newcommand{\EA}{\end{eqnarray}}

\newcommand{\vA}{\vec{A}}

\newcommand{\vAp}{\vA_{\rm p}}

\newcommand{\vB}{\vec{B}}

\newcommand{\vBp}{\vB_{\rm p}}

\newcommand{\vBj}{\vec{B}_{\rm j}}

\newcommand{\Cm}{C_{\rm m}}   
\newcommand{\Cr}{C_{\rm r}}   
\newcommand{\curlA}{\Nabla \times \vA}
\newcommand{\curlAp}{\Nabla \times \vAp}  

\newcommand{\divAp}{\Nabla \cdot \vAp}

\newcommand{\divB}{\Nabla \cdot \vB}

\newcommand{\dHdt}{{\rm d}H/{\rm d}t}
\newcommand{\dHmdt}{{\rm d}H_{\rm m}/{\rm d}t}
\newcommand{\fdHdt}{\frac{{\rm d}H}{{\rm d}t}}

\newcommand{\fdHmdt}{\frac{{\rm d}H_{\rm m}}{{\rm d}t}}
\newcommand{\E}{E_{\vol}}             
\newcommand{\Ep}{E_{\vol,\rm p}}      
\newcommand{\Ej}{E_{\vol,\rm ~j}}      

\newcommand{\vE}{\vec{E}}   
\newcommand{\epscm}{\epsilon_{\rm Cm}}
\newcommand{\epsH}{\epsilon_{\rm H}}
\newcommand{\FAAp}{F_{\rm AAp}}
\newcommand{\FBn}{F_{\rm Bn}}
\newcommand{\FVn}{F_{\rm Vn}}
\newcommand{\Fphi}{F_{\phi}}
\newcommand{\Ftot}{F_{\rm tot}}
\newcommand{\Hm}{H_{\rm m}}
\newcommand{\Hmix}{H_{\rm mix}}
\newcommand{\Hp}{H_{\rm p}}
\newcommand{\Hj}{H_{\rm j}}
\newcommand{\Hpj}{H_{\rm pj}}

\newcommand{\vv}{\vec{v}}

\begin{document} 

\title{Testing magnetic helicity conservation in a solar-like active event}
\author{
E.~Pariat\inst{1} 
\and  
G.~Valori\inst{2}  
\and
P.~D\'emoulin\inst{1}  
\and  
K.~Dalmasse \inst{3}\fnmsep  \inst{1}
       }
\institute{
LESIA, Observatoire de Paris, PSL Research University, CNRS, Sorbonne Universit\'es, UPMC Univ. Paris 06, Univ. Paris Diderot, Sorbonne Paris Cit\'e, 5 place Jules Janssen, 92195 Meudon, France \email{etienne.pariat@obspm.fr}\\
         \and
UCL-Mullard Space Science Laboratory, Holmbury St. Mary, Dorking, Surrey, RH5 6NT, UK     \\
         \and
CISL/HAO, National Center for Atmospheric Research, P.O. Box 3000, Boulder, CO 80307-3000, USA \\
}

\date{Received ***; accepted ***}

   \abstract
{
Magnetic helicity has the remarkable property of being a conserved quantity of ideal magnetohydrodynamics (MHD).  Therefore, it could be used as an effective tracer of the magnetic field evolution of magnetised plasmas.
}
{
Theoretical estimations indicate that magnetic helicity is also essentially conserved
with non-ideal MHD processes, e.g. magnetic reconnection. This conjecture has however been barely tested, either experimentally or numerically. Thanks to recent advances in magnetic helicity estimation methods, it is now possible to test numerically its dissipation level in general three-dimensional datasets.
}
{We first revisit the general formulation of the temporal variation of relative magnetic helicity on a fully bounded volume when no hypothesis on the gauge is made.
We  introduce a method to precisely estimate its dissipation independently of the type of non-ideal MHD processes occurring. In a solar-like eruptive 
event simulation, using different gauges, we compare its estimation in a finite volume with its time-integrated flux through the boundaries, hence testing the conservation and dissipation of helicity.
}
{We provide an upper bound of the real dissipation of magnetic helicity: It is quasi-null during the quasi-ideal MHD phase. 
Even when magnetic reconnection is acting the relative dissipation of magnetic helicity is also very small ($<2.2\%$), in particular compared to the relative dissipation of magnetic energy ($>30$ times larger). We finally illustrate how the helicity-flux terms involving velocity components are gauge dependent, hence limiting their physical meaning.  
}
{Our study paves the way for more extended and diverse tests of the magnetic helicity conservation properties.  Our study confirms the central role that helicity can play in the study of MHD plasmas. For instance, the conservation of helicity can be used to track the evolution of solar magnetic fields, from its formation in the solar interior until their detection as magnetic cloud in the interplanetary space. 
}
    \keywords{Magnetic fields, Methods: numerical, Sun: surface magnetism, Sun: corona}

   \maketitle

\section{Introduction} \label{s:intro}

In physics, conservation principle have driven the understanding of observed phenomena. Exact and even approximately conserved 
quantities have allowed to better describe and predict the behaviour of physical systems. Conservation laws state that, for an isolated system,
 a particular measurable scalar quantity does not change as the system evolves. A corollary is that for a non isolated system, a conserved 
 scalar quantity only evolves thanks to the flux of that quantity through the studied system boundaries. Given a physical paradigm, a physical 
 quantity may not be conserved if source or dissipation terms exist. 
 
In the magnetohydrodynamics (MHD) framework, a quantity has received increasing attention for its conservation property: 
magnetic helicity \citep{Elsasser56}.  Magnetic helicity quantitatively describes the geometrical degree of twist, shear, or more generally, 
knottedness of magnetic field lines \citep{Moffatt69}. In ideal MHD, where magnetic field can be described as the collection of individual magnetic
field lines, magnetic helicity is a strictly conserved quantity \citep{Woltjer58} as no dissipation, nor creation, of helicity is permitted 
since magnetic field line cannot reconnect.

In his seminal work, \citet{Taylor74} conjectured that even in non-ideal MHD, the dissipation of magnetic helicity should be relatively weak, 
and hence, that magnetic helicity should be conserved. This could be theoretically explained by the inverse-cascade property of magnetic 
helicity: in turbulent medium, helicity unlike magnetic energy, tends to cascade towards the larger spatial scales, 
thus avoiding dissipation at smaller scales \citep{Frisch75,Pouquet76}. This cascade has been observed in numerical simulations 
\citep{Alexakis06,Mininni07} as well as in laboratory experiments \citep{Ji95}. In the case of resistive MHD, \citet{Berger84} derived an upper limit on 
the amount of magnetic helicity that could be dissipated through constant resistivity. He showed that the typical helicity dissipation time in the solar 
corona was far exceeding the one for magnetic energy dissipation. 

From the Taylor's conjecture on helicity conservation have been derived multiple important consequences for the dynamics of plasma 
systems.  Based on helicity conservation, \citet{Taylor74} predicted that relaxing MHD systems should reach a linear force free state.  
This prediction, which was verified to different degrees, has allowed to understand the dynamics of plasma in several laboratory 
experiments \citep{Taylor86,Prager99,Ji99,Yamada99}. The concept of Taylor relaxation has also driven 
theoretical models of solar coronal heating \citep{Heyvaerts84}. The importance of magnetic helicity conservation has been further raised 
for magnetic field dynamos \citep{Brandenburg05,Blackman14}. Magnetic helicity is also strongly impacting the energy budget during reconnection  
events \citep{Linton01,Linton02,DelSordo10}, and models of eruption based on magnetic helicity annihilation have been developed \citep{Kusano04}. 
The conservation of magnetic helicity has been suggested as the core reason of the existence of Coronal Mass Ejections (CMEs), 
the latter being the mean for the Sun to expel its excess magnetic helicity  \citep{Rust94,Low96}. Because of this hypothesis, important 
efforts to estimate the magnetic helicity in the solar coronal have been carried over the last decades \citep{Demoulin07,Demoulin09}.

Despite its potential importance, it is surprising to note that Taylor's conjecture on helicity conservation has been barely tested. Several 
experimental and numerical works have focused on testing Taylor's prediction that relaxed systems should be linear force free  
\citep{Taylor86,Prager99,Ji99,Yamada99,Antiochos99}. However, the fact that the system may not reach a complete relaxed linear force free state, 
does not mean that helicity is not well conserved. It simply means that the dynamics of the relaxation does not allow the full redistribution of helicity 
towards the largest available scales. Looking at the long term evolution of turbulent MHD systems, magnetic helicity has been found to decay slower
than magnetic energy \citep{Biskamp99,Christensson05,Candelaresi12}. These studies are however based on the estimation of the 
helicity density which is not generally gauge invariant. The periodic boundaries, the long term and large spatial scales involved do not 
allow an estimation of helicity dissipation when a "singular" non-ideal event is occurring.   

Direct tests on magnetic helicity conservation and dissipation have actually been limited because of the inherent 
difficulties to measure that very quantity. Laboratory experiments and observations requires the measurement of the full 3D 
distribution of the magnetic field. Most laboratory experiments therefore make assumptions on the symmetries of the system to limit the sampling of the 
data, hence limiting measurement precision.  
\citet{Ji95} have measured a $1.3-5.1 \%$ decay of helicity (relatively to a $4-10\%$ decrease of energy) during a sawtooth relaxation 
in a reversed-field pinch experiment.  \citet{Heidbrink00} have estimated that helicity was conserved  within $1\%$ during a sawtooth crash. Other 
experiments have tried to measure helicity conservation by comparing the helicity in the system with its theoretically injected amount 
\citep{Barnes86,Gray10}. While they found results agreeing within $10-20\%$, the experimental conditions limit the precision on the measure of 
injected helicity \citep{Stallard03}. In numerical experiments of flux tubes reconnection, using triple periodic boundaries, 
\citet{Linton05} have found that the loss of helicity was ranging between $6 \%$ and $53 \%$ depending on the Lundquist number, 
hence primarily due to diffusion rather than magnetic reconnection.

For non-periodic systems where the magnetic field is threading the domain boundary (as in most natural cases), the very definition of 
magnetic helicity is not gauge invariant (cf. \sect{theoryH}), and a modified definition of magnetic helicity, the relative magnetic helicity, had to be 
introduced \citep[][see \sect{theoryHrel}]{BergerField84}. Only very recently, practical methods that allow to generally computed 
relative magnetic helicity have been published \citep{Rudenko11,Thalmann11,Valori12,YangS13}. These methods are now 
opening the possibilities to carefully study helicity of 3D datasets such as the ones frequently used for natural plasmas. 
\citet{ZhaoL15} and \citet{Knizhnik15} have observed a good helicity conservation but have not quantified it. \citet{YangS13} have 
measured the helicity conservation in a numerical simulation by comparing the relative helicity flux with the variation of helicity in the domain. 
They found that during the quasi-ideal evolution of the system the helicity was conserved within $3\%$, with higher dissipation values when non-ideal 
MHD effects become important. However, in previous studies, the relative helicity flux definition used may not be fully consistent 
with the choice of gauge used to compute the volume helicity (see \sect{theoryHrel}).

In the present manuscript, we will further push the test on the conservation of magnetic helicity. We will derive a generalized analytical formula 
for the flux of relative magnetic helicity (\sect{theoryHrelvar}) without taking any assumption on the gauges of the studied and the reference fields. 
We will discuss whether relative magnetic helicity can be considered as a conserved quantity in the classical sense, that is, whether its variations 
can be described solely as a flux through the boundary.  The general method that we will employ (see \sect{Methodology}) is 
based on the comparison of the evolution of the relative magnetic helicity with its flux through the boundaries, and will 
be applied to a numerical simulation of solar active-like events. In Sections~\ref{s:DeVore} \& \ref{s:DeVore-Coulomb}, 
using different gauges we will constraint the level of conservation of relative magnetic helicity, and study the amount of dissipated magnetic helicity.
This is further extended in Appendix~\ref{s:GenDev-Coul} with another selection of gauges. We will finally conclude in \sect{Conclusion}

\section{Magnetic helicity and its time variation} \label{s:theory}

\subsection{Magnetic helicity} \label{s:theoryH}

Magnetic helicity is defined as
\BE 
\label{eq:defH}
\Hm = \int_{\vol} \vA \cdot \vB \dV \, ,
\EE
where $\vB$ is the magnetic field studied over a fixed volume $\vol$.  $\vol$ is here a fully closed\footnote{We do not consider the particular cases of infinite volumes which require specific hypothesis on the behaviour of the quantities at large distance.} volume bounded by the surface $\surf$ . The vector potential, $\vA$ of $\vB$ classically verifies:
$\curlA = \vB$ as $\divB =0$ \citep[only approximately true in numerical computations,][]{Valori13}. 
The magnetic field $\vB$ is gauge invariant, meaning that it is unchanged by transformations $\vA \rightarrow \vA+\Nabla \psi$, where $\psi$ is any 
sufficiently regular, scalar function of space and time. Since $\vA$ is not uniquely defined in general, 
magnetic helicity requires additional constraints in order to be well-defined. In particular,  
$\Hm$ is a gauge-invariant quantity provided that $\vol$ is a magnetic volume, i.e. that the magnetic field is tangent at any 
point of the surface boundary $\surf$ of $\vol$: $(\vB \cdot \dS)\left|_\surf\right.=0$, at any time.

Assuming that the volume $\vol$ is fixed, with $\surf$ a flux surface ensuring gauge invariance, the temporal variation of magnetic helicity is derived by direct differentiation of \eq{defH}:
\BE 
\label{eq:dHsdt1}
\fdHmdt = \int_{\vol} \pardiff{\vA}{t} \cdot \vB \dV 
        + \int_{\vol} \vA \cdot \pardiff{\vB}{t} \dV \, , 
\EE
where each of the two integrals on the right hand side are well defined, since they are independently gauge invariant. Given that $\Nabla \cdot (\vA \times \partial \vA/ \partial t)= \partial \vA/ \partial t \cdot \curlA - \vA \cdot \partial (\curlA)/ \partial t$ and using the Gauss divergence theorem, one obtains:
\BE 
\nonumber
\fdHmdt  = \int_{\surf} \left(\vA \times \pardiff{\vA}{t} \right) \cdot \dS  
        + 2\int_{\vol} \vA \cdot \pardiff{\vB}{t} \dV  \, .
\EE
Here, $ \dS$ is the elementary surface vector, directed outside of the domain  $\vol$. Using the Faraday's law of induction, one derives:
\BE 
\label{eq:dHsdt3}
\fdHmdt 
  =  \int_{\surf} \left(\vA \times \pardiff{\vA}{t} \right) \cdot  \dS  
   - 2\int_{\vol} \vA \cdot \Nabla \times \vE \dV   \, , 
\EE
Using again the Gauss divergence theorem, one finds that the temporal variation of magnetic helicity is composed of three 
independently gauge-invariant terms, \ie a volume dissipative term and two helicity flux terms on the surface of $\vol$, such that:
\BA 
\fdHmdt 
        &=& \left.\fdHmdt \right |_{\rm diss} +  F_{m,B} 
           + F_{m,A}\quad \text{with}   \label{eq:dHsdt4} \\
\left. \fdHmdt \right |_{\rm diss} 
        &=& - 2\int_{\vol} \vE \cdot \vB \dV  \label{eq:Hdiss}\\
F_{m,B} &=& 2\int_{\surf} (\vA \times \vE) \cdot \dS 
        \label{eq:FmB}\\
F_{m,A} &=& \int_{\surf} \left(\vA \times \pardiff{\vA}{t} \right) \cdot \dS 
        \label{eq:FmA}
 \EA

In ideal MHD, where $\vE=-\vv \times \vB$, the volume term is null. For an isolated system, magnetic helicity is thus conserved in the classical sense
 since its variations are null (cf. \sect{intro}). Variations of $\Hm$ can only originate from advection of helicity through the boundaries of $\vol$. 
The $\dHmdt|_{\rm diss}$ term corresponds to the dissipation of magnetic helicity of the studied magnetic field in $\vol$. 
\citet{Taylor74} conjectured that this term is relatively small even when non-ideal MHD processes are developing, \eg when magnetic reconnection 
is present (cf. \sect{intro}).

However, because of the gauge invariance requirement which imposes that $\surf$ must be a flux surface, magnetic helicity appears as a 
quantity of limited practical use. In most studied systems, magnetic field is threading the surface $\surf$ and the condition
$(\vB \cdot \dS)\left|_\surf\right.=0$ is not fulfilled. In their seminal paper, \citet{BergerField84} have introduced the concept of relative magnetic helicity: 
a gauge-invariant quantity which preserve essential properties of magnetic helicity while allowing a non null normal component of the 
field $\vB$ through the surface of the studied domain.

\subsection{Relative Magnetic helicity}\label{s:theoryHrel}

In their initial work, \citet{BergerField84} gave a first definition of the relative magnetic helicity as the difference between the helicity of the studied field $\vB$ and the helicity of a reference field $\vB_0$ having the same distribution than $\vB$ of the normal component along the surface: $((\vB_0-\vB) 
\cdot \dS)\left|_\surf\right.=0$. 

While the definition allows for any field to be used as the reference field, the potential field $\vBp$ is frequently used as a reference field in the literature. Since $\Nabla \times \vBp=0$, the potential field can be derived from a scalar function $\phi$:
\BE 
\label{eq:phi}
\vBp = \Nabla \phi
\EE  
where the scalar potential $\phi$ is the solution of the Laplace equation $\triangle \phi =0$ derived from $\Nabla \cdot \vBp=0$.
Given the distribution of the normal component on the surface $\vB \cdot \dS\left|_\surf\right. = \partial \phi / \partial n$ of the studied domain, at any instant there is a unique potential field which satisfies the following condition on the whole boundary of the volume considered: 
\BE 
\label{eq:condrelhgauge}
(\vBp\cdot \dS)\left|_\surf\right. = (\vB \cdot \dS)\left|_\surf\right. 
\EE  
Under the above assumptions, the potential field has the lowest possible energy for the given distribution of $\vB$ on $\surf$ 
\citep[e.g. Eq.(2) of][]{Valori12}. 
In the following we will also use the potential field as our reference field. 
 
A second gauge-independent definition for relative magnetic helicity has been given by \citet{Finn85}, definition used from here on in this article:
\BE 
\label{eq:defHrel}
H = \int_{\vol} (\vA+\vAp ) \cdot (\vB-\vBp ) \dV \, .
\EE
with $\vAp$ the potential vector of the potential field $\vBp=\curlAp$. 
Not only is $H$ gauge invariant, but the gauges of $\vA$ and $\vAp$ are independent of each others, \ie for any set of sufficiently-regular 
scalar functions $(\psi;\psi_p)$, $H$ will be unchanged by the gauges transformation $(\vA;\vAp) \rightarrow (\vA+\Nabla \psi; \vAp+\Nabla \psi_p)$.
Let us note that $\vAp$ and $\phi$ correspond to two distinct 
solutions of the Helmholtz's theorem, i.e. two distinct non-incompatible decompositions of $\vBp$. 

The relative helicity in \eq{defHrel} can first be decomposed in a contribution due only to $\vB$, \eq{defH}, one only to $\vBp$, and a mixed term:
\BA 
H     & = & \Hm - \Hp + \Hmix \quad \text{with} \label{eq:HrelDecomp} \\
\Hp   & = & \int_{\vol} \vAp \cdot \vBp \dV \\  \label{eq:defHp}
\Hmix & = & \int_{\vol} (\vAp \cdot \vB - \vA \cdot \vBp) \dV = \int_{\surf} (\vA \times \vAp) \cdot \dS \label{eq:defHmix} 
\EA
Let us note that this decomposition is only formal in the sense that each term is actually gauge dependent and only their sum is gauge invariant.


Relative magnetic helicity, as defined in \eq{defHrel}, is equal to the difference between the helicity of the field $\vB$ and the helicity of its 
potential field ($H = \Hm(\vB)-\Hm(\vBp)$) only if $\Hmix$ cancels. Relative helicity is in general not a simple difference of helicity, as for 
relative energy.
A sufficient (but not necessary) condition that ensures the nullity of the mixed term is that $\vA$ and $\vAp$ have the same transverse component 
on the surface, i.e.:
\BE 
\label{eq:condrelhdiff}
\vA \times \dS|_\surf = \vAp \times \dS|_\surf 
\EE
This condition automatically enforces the condition of \eq{condrelhgauge} on the normal field components. However it 
imposes to link the choice of the gauge of $\vA$ with that of $\vAp$. The original definition of \citet{BergerField84} corresponds to a quantity which 
is less general than the one given by \citet{Finn85}. It is only gauge invariant for particular sets of transformation: 
$(\vA;\vAp) \rightarrow (\vA+\Nabla \psi; \vAp+\Nabla \psi_p)$ . 
  
 Another possible decomposition of relative magnetic helicity from \eq{defHrel} is \citep{Berger03}:
 \BA 
H          &=& \Hj + 2\Hpj  \quad \text{with} 
               \label{eq:HrelDecomp2}\\
\Hj &=& \int_{\vol} (\vA - \vAp)  \cdot (\vB-\vBp) \dV 
               \label{eq:defHj}\\
\Hpj &=& \int_{\vol} \vAp \cdot (\vB-\vBp) \dV 
               \label{eq:defHpj}
\EA
where $\Hj$ is the classical magnetic helicity of the non-potential, or current carrying, component of the magnetic field, $\vBj=\vB-\vBp$ and $\Hpj$ is the mutual helicity between $\vBp$ and $\vBj$. The field $\vBj$ is contained within the volume $\vol$ so it is also called the closed field part of $\vB$. Because of \eq{condrelhgauge}, not only $H$, but also both $\Hj$ and $\Hpj$ are independently gauge invariant. 

\subsection{Relative magnetic helicity variation}\label{s:theoryHrelvar}

Assuming a fixed domain $\vol$, we can differentiate \eq{defHrel} in time in order to study the time variations of relative helicity: 
\BE 
\label{eq:Hvar1}
\begin{split} 
\fdHdt 
 &=\int_{\vol} \pardiff{(\vA+\vAp)}{t} \cdot \Nabla \times (\vA-\vAp) \dV  \\
 &+\int_{\vol} (\vA+\vAp) \cdot \pardiff{(\vB-\vBp)}{t} \dV       
\end{split}
\EE
Using the Gauss divergence theorem one obtains:
\BE 
\nonumber
\begin{split} 
\fdHdt  
  &= \int_{\surf} \left((\vA-\vAp) \times \pardiff{(\vA+\vAp)}{t} \right) \cdot \dS \\
  &+ \int_{\vol} (\vA-\vAp) \cdot \pardiff{(\vB+\vBp)}{t} \dV \\
  &+ \int_{\vol} (\vA+\vAp) \cdot \pardiff{(\vB-\vBp)}{t} \dV        
\end{split}
\EE
Combining the second and third terms, one finds the following synthetic decomposition of the helicity variation in three terms:  
\BE 
\label{eq:Hvar3}
\begin{split} 
     \fdHdt 
  =& 2\int_{\vol} \vA \cdot \pardiff{\vB}{t} \dV  
    + \int_{\surf} \left((\vA-\vAp) \times \pardiff{(\vA+\vAp)}{t} \right) \cdot \dS \\
  -& 2\int_{\vol} \vAp \cdot \pardiff{\vBp}{t} \dV 
\end{split}
\EE
This decomposition is only formal. Indeed, as for the decomposition of relative helicity of \eq{HrelDecomp}, none of these three terms are independently gauge invariant and only their sum is. 
The third term can be further decomposed using the scalar potential $\phi$, \eq{phi}, and the Gauss divergence theorem:
\BE 
\label{eq:Hvarbp}
\begin{split}
 \left.\fdHdt\right|_{Bp}  
 &= - 2\int_{\vol} \vAp \cdot \pardiff{\vBp}{t} \dV   
  = -2\int_{\vol} \vAp \cdot \Nabla \left(\pardiff{\phi}{t}\right) \dV \\
 &= -2\int_{\surf} \pardiff{\phi}{t} \vAp \cdot  \dS +2 \int_{\vol}\pardiff{\phi}{t} \divAp \dV 
\end{split}
\EE
Using the Faraday law and the Gauss divergence theorem we also obtain: 
\BE 
\label{eq:Hvarb}
\int_{\vol} \vA \cdot \pardiff{\vB}{t} \dV  
  = -\int_{\surf} (\vE \times \vA) \cdot \dS  
    -\int_{\vol} \vB \cdot \vE \dV 
\EE
Assuming that at the boundary the evolution of the system is ideal: $\vE\left|_\surf\right.= (-\vv \times \vB)\left|_\surf\right.$ the surface flux 
can be written as \citep[e.g.][]{BergerField84}:
\BE 
\label{eq:HvarExA}
  -\int_{\surf} (\vE \times \vA) \cdot \dS  
= -\int_{\surf} (\vB \cdot \vA) \vv \cdot \dS
  +\int_{\surf} (\vv \cdot \vA) \vB \cdot \dS  
\EE
Note that if the evolution of the system was not ideal at the boundary, an additional flux term 
depending on $\vE_{\text{non ideal}} \times \vA$,  could be added, with $\vE_{\text{non ideal}}$ the non ideal part of the electric field. 
In the study presented here this term will de facto be estimated but assumed to be measured as a volume dissipation term.

Including \eqss{Hvarbp}{HvarExA} in \eq{Hvar3}, the variation of magnetic helicity can thus be decomposed as: 
\BE 
\label{eq:Hvar4}
\fdHdt = \left. \fdHdt \right |_{\rm diss} + \left. \fdHdt \right|_{\rm Bp,var} 
       + \FVn + \FBn + \FAAp + \Fphi
\EE
   with
\BA 
\left.\fdHdt\right |_{\rm diss}
      &=& -2\int_{\vol} \vE \cdot \vB \dV               \label{eq:defdHdiss} \\
\left.\fdHdt\right|_{\rm Bp,var} 
      &=& \phantom{-} 2\int_{\vol}\pardiff{\phi}{t} \divAp \dV     \label{eq:defdHbpvar} \\
\FVn  &=& -2\int_{\surf} (\vB \cdot \vA) \vv \cdot  \dS \label{eq:defFVn} \\ 
\FBn  &=& \phantom{-} 2\int_{\surf} (\vv \cdot \vA) \vB \cdot \dS  \label{eq:defFBn}\\  
\FAAp &=& \phantom{-2}  \int_{\surf} \left((\vA-\vAp) \times \pardiff{(\vA+\vAp)}{t}\right)
                 \cdot \dS                              \label{eq:defFAAp}\\ 
\Fphi &=& -2\int_{\surf} \pardiff{\phi}{t} \vAp \cdot \dS
                                                        \label{eq:defFphi}
\EA
\
The $\dHdt |_{\rm diss}$ term is a volume term which corresponds to the actual dissipation of magnetic helicity of the studied magnetic 
field (Eq.~\ref{eq:Hdiss}). The $\dHdt |_{\rm Bp,var}$ term, which, despite being a volume term, is not a dissipation, traces a change in the 
helicity of the potential field. 
As $\vB$ is evolving, its distribution at the boundary implies a changing $\vBp$ (Eq.~\ref{eq:condrelhgauge}). The helicity of the potential field, 
not necessarily null, is therefore evolving in time. More precisely, the potential field is defined only in terms of its boundary values. However, this is not 
true in general for the vector potential of the potential field, because of the gauge freedom. Hence, in general, the helicity of the potential field cannot 
be expressed as a function of boundary values only, except for the particular case of a vector potential without sources or sinks in $\vol$, \ie  when the 
Coulomb gauge is used. Therefore, the time variation of the helicity of the potential field necessarily contains both volume and flux contributions.

All the other terms, are flux terms that correspond to the transfer of helicity through the surface boundary $\surf$. The $\FVn$ and $\FBn$ are sometimes called the "emergence" and "shear" terms, but such a characterization can be misleading as their contributions depend on the gauge selected for $\vA$.  The $\FAAp$ term is related to a cross contribution of $\vA$ and $\vAp$. Finally 
$\Fphi$ correspond to a flux of the helicity of the potential field.

The $\dHdt |_{\rm diss}$ term is the only term of the decomposition which is gauge invariant.  All the other terms are not independently gauge invariant.  
It means that the relative intensity of these terms will be different for different gauges.  
Combined, they produce the same, gauge-invariant value of $\dHdt$. 
We will study the dependence of the above decomposition on the chosen gauge in  \sect{DeVore-Coulomb}. 
We should also note that the total flux $\Ftot$ of relative helicity,
\BE 
\label{eq:defFtot}
 \Ftot = \FVn + \FBn + \FAAp + \Fphi \ ,
\EE
is only gauge invariant for $\vA$ but not for $\vAp$.

Moreover, unlike $\dHdt |_{\rm diss}$, $\dHdt |_{\rm Bp,var}$ is not a priori null in ideal MHD. 
It implies that $\dHdt$ cannot be written in a classical conservative form since $\dHdt |_{\rm Bp,var}$ cannot be strictly written as a flux term.
Therefore relative magnetic helicity is not a priori a conserved quantity of MHD in the classical sense: its variation in $\vol$ may not solely come 
from a flux of relative helicity through the boundary. Relative magnetic helicity may not be conserved even 
if $\dHdt |_{\rm diss}$ is small. The conservation of relative helicity and the dissipation of magnetic helicity are thus two distinct problems: the relative intensity of these terms will vary, depending on which gauge is employed as we will illustrate in \sect{DeVore-Coulomb}. 

\subsection{Relative Magnetic helicity variation with specific gauge conditions} \label{s:theoryHrelvarSpec}

While the variation of magnetic helicity can be generally described by \eq{Hvar4} for any gauge,  the choice of some specific additional constraint on the gauge allows to simplify the expression of $\dHdt$ and possibly its computation.

\subsubsection{The $\vA|_\surf =  \vAp|_\surf$ condition}
 \label{s:theoryHrelvarSpec1}
 
We note that with the specific condition
\BE 
\label{eq:CondA=Ap}
\vA|_\surf =  \vAp|_\surf \ ,
\EE
the condition of Eq. (\ref{eq:condrelhdiff}) is necessarily satisfied thus $\FAAp=0$ and 
the terms $\FVn$ and $\FBn$ can be expressed only in terms of $\vAp$.
Thus, the helicity variation, \eq{Hvar4}, simplifies as:
\BE 
\begin{split} \label{eq:HvarcondAAp}
\left. \fdHdt\right |_{\rm Cond. (\ref{eq:CondA=Ap})}  =  
&-2\int_{\vol} \vE \cdot \vB \dV 
 +2\int_{\vol}\pardiff{\phi}{t} \divAp \dV \\
&-2\int_{\surf} (\vB \cdot \vAp) \vv \cdot  \dS  
 +2\int_{\surf} (\vv \cdot \vAp) \vB \cdot \dS \\
&-2\int_{\surf} \pardiff{\phi}{t} \vAp \cdot \dS
\end{split}
\EE

We note that in this derivation the vector potential $\vA$ is absent. The condition (\ref{eq:CondA=Ap}) 
allows to get rid of the need of computing $\vA$ to estimate the helicity variations. 
However, to derive $H$ from \eq{defHrel} both $\vA$ and $\vAp$ must be computed with gauges coupled with \eq{CondA=Ap}. 
Then, one must strictly control that this condition is enforced all over the surface of the studied system. This can actually be numerically challenging. 

In the present manuscript, in order to determine the helicity dissipation, we will compare time integrated helicity flux with direct 
helicity measurements. As our numerical method does not allow us to enforce \eq{CondA=Ap}, we cannot use the 
simplified \eq{HvarcondAAp} to compute the helicity variation.

\subsubsection{Coulomb gauge for potential field: $\divAp = 0$}
 \label{s:theoryHrelvarSpec2}
 
If to determine the potential field one decides to choose the Coulomb gauge, 
\BE 
\label{eq:Coulomb}
\divAp = 0\ ,
\EE
 then there is no volume variation of the helicity of the potential field $\dHdt |_{\rm Bp,var}=0$. The helicity variation can be reduced to the simplified form:
\BE 
\label{eq:VarHCoulomb}
\left.\fdHdt\right |_{\rm Cond.  (\ref{eq:Coulomb})}  = -2\int_{\vol} \vB \cdot \vE \dV + \Ftot
\EE 
Using the coulomb gauge for the potential field one observes that the variation of the relative magnetic helicity are given by a flux of helicity 
through the boundary and the dissipation term $\dHdt |_{\rm diss}$. Relative magnetic helicity to a reference field expressed in the Coulomb gauge 
can therefore be written with a classical conservative equation.

\subsubsection{Boundary null Coulomb gauge}
 \label{s:theoryHrelvarSpec3}
 
The condition (\ref{eq:Coulomb}) does not enforce a unique solution for $\vAp$.
It is possible to further constrain the Coulomb gauge if the vector potential $\vAp$ satisfies the additional boundary condition: 
\BE 
\label{eq:condgaugeap}
\vAp \cdot \dS|_\surf = 0 \ .
\EE

With this further constraint the flux of helicity of the potential field is null, $\Fphi=0$. Then, with conditions (\ref{eq:Coulomb}) and (\ref{eq:condgaugeap}), the helicity variation thus reduces to the form:
\BE 
\label{eq:VarHCoulomb2}
\begin{split}
\left.\fdHdt\right |_{\rm Cond.  (\ref{eq:condgaugeap})}=
&- 2\int_{\vol} \vE \cdot \vB \dV + \int_{\surf} \Bigg( 2(\vv \cdot \vA) \vB \\
&- 2(\vB \cdot \vA) \vv + (\vA-\vAp) \times \pardiff{(\vA+\vAp)}{t} \Bigg) \cdot \dS
\end{split}
\EE 

\subsubsection{Simplified helicity flux}
 \label{s:theoryHrelvarSpec4}
 
Including the three conditions of previous sub-sections, \ie: 
\BE 
\label{eq:condgaugeorgy}
\left\{\begin{array}{rl} 
\vA|_\surf            &=\;\; \vAp|_\surf \\
\divAp                &=\;\; 0 \\
\vAp \cdot \dS|_\surf &=\;\; 0 \ .
\end{array} \right.
\EE
we obtain the well known expression for the simplified helicity flux \citep[e.g.][]{BergerField84,Pariat05}:
\BE 
\label{eq:VarHCoulomb3}
\begin{split}
\left.\fdHdt\right |_{\rm Cond.\ (\ref{eq:condgaugeorgy})} =
&- 2\int_{\vol} \vE \cdot \vB \dV \\
&+ 2\int_{\surf} ((\vv \cdot \vAp) \vB - (\vB \cdot \vAp) \vv) \cdot \dS 
\end{split}
\EE 

With this set of conditions, the flux terms $\FVn$ and $\FBn$ are fully fixed. However, one should remind that these terms remain gauge dependent: using a gauge where the conditions of \eq{condgaugeorgy} are not fully enforced would lead to a different $\vAp$ and $\vA$, consequently, to a different distribution of helicity flux between  $\FVn$, $\FBn$ and other terms. It is therefore theoretically incorrect to study them independently.  

Equation (\ref{eq:VarHCoulomb3}) is the classical formulation for the helicity flux that has been derived by \citet{BergerField84} for an infinite plane. 
However in the case of a 3D cubic domain this formulation is only valid if all the conditions of \eq{condgaugeorgy} are satisfied.
While \citet{YangS13} have used \eq{VarHCoulomb3} to compute the helicity flux, it remains to be determined if all conditions (\ref{eq:condgaugeorgy}) are fulfilled when computing the volume helicity. 
Their relatively high level of non-conservation (3\%) in the ideal phase of the evolution of their system may be related to this discrepancy.
Finally, while conditions of Eq. (\ref{eq:condgaugeorgy}) can drastically simplify the estimation of the helicity variation, they strongly 
constraint the numerical implementation of $\vA$ and $\vAp$. Fast, precise and practical numerical computation of the 
vector potentials may require a different choice a gauge.

\section{Methodology} \label{s:Methodology}

The present section describes the methodology employed in our numerical experiments. In \sect{Mestimators} we present the general method used to estimate the relative magnetic helicity conservation and the magnetic helicity dissipation. Then in \sect{Mhelicomp} we describe how we practically compute the volume helicity and its flux. Finally, in \sect{Msimulations} we present the numerical data set considered for the helicity conservation tests.

\begin{figure*}[ht]
  \setlength{\imsize}{0.99\textwidth}
  \includegraphics[width=\imsize,clip=true]{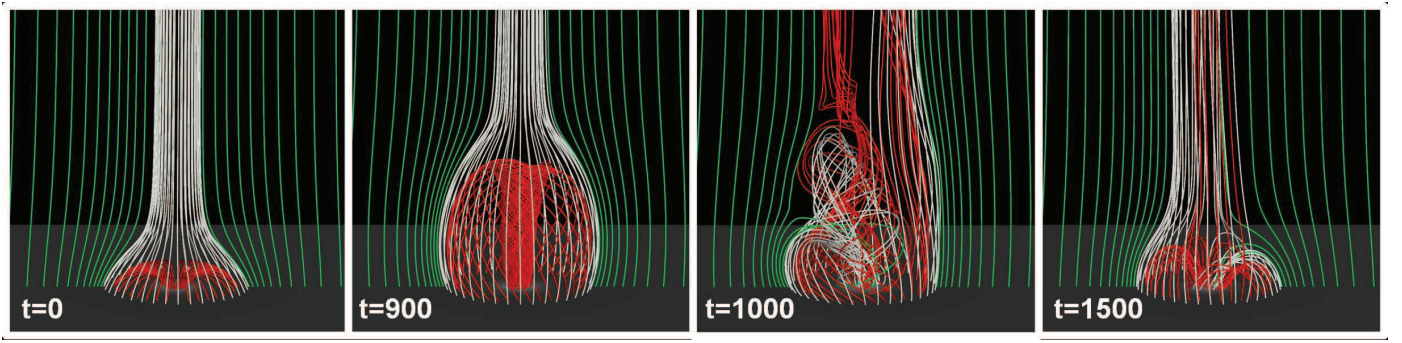}
  \caption{Snapshots of the magnetic field evolution during the generation of the jet. The red field lines are initially closed. The green and white field lines are initially open. All the field lines are plotted from fixed footpoints. The red and white field lines are regularly plotted along a circle of constant radius while the green field lines are plotted along the $x$ axis. At $t=900$ the system is in its pre-eruption stage. It is close to the maximum of energy and helicity. All the helicity is stored in the close domain. At $t=1000$ the system is erupting. Numerous field lines have changed of connectivity as one observe open red field lines and closed white field lines. Helicity is ejected upward along newly open reconnected field lines.  At $t=1500$ the system is slowly relaxing to its final stage.}
  \label{Fig:jetsim}
\end{figure*}

\subsection{Estimators of magnetic helicity conservation}  \label{s:Mestimators}

Following \citet{YangS13}, we will compute the volume variation along with the flux of magnetic helicity.
From two successive outputs of the studied MHD system,  
corresponding to two instant $\tau$ and $\tau'$, separated by a time interval $\Delta t$ we will 
directly compute their respective total helicity $H_\vol(\tau)$ and $H_\vol(\tau')$ in the volume $\vol$ using the method of \sect{Mhelicomp}, 
and then, the helicity variation rate between these 2 instants:
$\Delta H_\vol/\Delta t = (H_\vol(\tau')-H_\vol(\tau))/\Delta t$. We will simultaneously estimate the different sources of fluxes 
$F_{\#}$ through the surface of the domain, with $\#$ the different contribution to the total flux in \eq{defFtot}. 

We will also time integrate the helicity fluxes at the boundary $H_\surf$:
\BA 
H_\surf (t)     &=& \int_0^t  \Ftot(\tau) ~\rmd \tau  \label{eq:defHsurftot} \\
H_{\surf,\#}(t) &=& \int_0^t F_{\#}(\tau) \rmd \tau  \label{eq:defHsurf}
\EA
with $\#$ the different contribution to the total flux of \eq{defFtot}. 

One should note that two very different methods are used to compute $H_\vol$ and $H_\surf$. To derive $H_\vol$ (and $\Delta H_\vol/\Delta t$) 
from \eq{defHrel}, only three components of the magnetic field $\vB$ in the whole domain $\vol$  are needed: $\vA,\vBp,\vAp$ are derived from $\vB$ 
(cf \sect{Mhelicomp}). For $H_\surf$ (and $\Ftot$), only data along the boundary $\surf$ are required. Furthermore, 
the helicity flux estimation requires the knowledge of the three components of the velocity field on $\surf$ (to compute $\FVn$ and $\FBn$). 
These quantities are not used to compute $H_\vol$. The methods of estimations of $H_\vol$ and $H_\surf$ are thus fully independent.

A physical quantity is classically said to be conserved when its time variation in a given domain is equal to its flux through the boundary of the 
domain (cf. \sect{intro}). To study the twofold problem of the conservation of 
relative magnetic helicity and the dissipation of magnetic helicity, we will use 
two quantities, $\Cr$ and $\Cm$, respectively.

For relative magnetic helicity to be perfectly conserved one should have $H_\vol=H_\surf$, or $dH/dt$ must be 
equal to $\Ftot$, i.e. the volume terms of the relative helicity variation should be null. By estimating, $\Cr$,   
\BE 
\label{eq:defCr}
\Cr =  \frac{\Delta H_\vol}{\Delta t} - \Ftot   \simeq \left. \fdHdt \right|_{\rm diss} 
    + \left. \fdHdt \right|_{\rm Bp,var} \ ,
\EE
we can determine the level of conservation of relative magnetic helicity. As already noted in \sect{theoryHrelvar} even if $\vB$ evolves within the ideal 
MHD paradigm, the relative magnetic helicity evolution cannot be written with a classical equation of conservation as the term in $\dHdt|_{\rm Bp,var}$ 
is not a flux integral through a boundary and generally does not vanish. 

Secondly, we aim to determine the dissipation of the magnetic helicity of the studied field (Eq.~\ref{eq:defdHdiss}), i.e. by estimating $\Cm$ equal to:
\BE 
\label{eq:defCm}
\Cm =  \frac{\Delta H_\vol}{\Delta t} - \Ftot - \left. \frac{\Delta H_\vol}{\Delta t} \right |_{\rm Bp,var}  \simeq   \left. \fdHdt \right|_{\rm diss}   
\EE
Our estimation of $\Cm$ is done independently of the estimation of $dH/dt|_{diss}$. Our method thus does not require the 
knowledge of the electric field $\vE$, which is a secondary quantity in most MHD problems. The dissipation is thus estimated in a way 
which is completely independent from the non-ideal process, e.g. the precise way magnetic reconnection is developing. 

As the estimators measure both physical helicity variations and numerical errors, we will use different 
non-dimensional criterions to quantify the level of helicity conservation and the precision of our measurements.
We will compute the relative accumulated helicity difference $\epsH$:
\BE 
\label{eq:defepsh}
\epsH (t) =  \frac{H_\vol (t)-H_\surf (t)}{H_{ref}} \simeq \frac{ \int_{\tau=0}^{t} \Cm ~\rmd \tau}{H_{ref}}
\EE
with $H_{ref}$ a normalizing reference helicity,  physically significant for the studied system (e.g. the maximum $H_\vol$ value in the studied interval).

At each instant, $\Cm$ expresses the rate of dissipation of helicity, \ie our numerical estimation of \eq{defdHdiss}. It can include 
artefact fluctuations due to the numerical precision of the flux estimation and the time derivation of $H_\vol$. In addition, as helicity is a 
signed quantity both positive and negative helicity can be generated by non-ideal effects. It may be relevant to know the time integrated 
absolute variation of helicity, as a function of time. Hence, we define another metric, $\epscm$,
\BE 
\label{eq:defepscm}
\epscm (t) =  \frac{ \int_{\tau=0}^{t} |\Cm | ~\rmd \tau}{\int_{\tau=0}^{t}|\Delta H_\vol/\Delta t| \rmd \tau},
\EE
where the absolute values guarantee that we take an upper limit of the dissipation. Along with $\epsH$, $\epscm$ allows us to evaluate the level of dissipation and the numerical precision of our measurements. The measure of low value of $\epscm$ and $\epsH$ can 
thus provide a clear demonstration of the level of conservation of relative magnetic helicity. For two methods having 
similar $\epsH$, a higher value of $\epscm$ would indicate measurements presenting larger numerical errors.

\begin{figure}[ht]
  \setlength{\imsize}{0.45\textwidth}
  \sidecaption    
  \includegraphics[width=\imsize,clip=true]{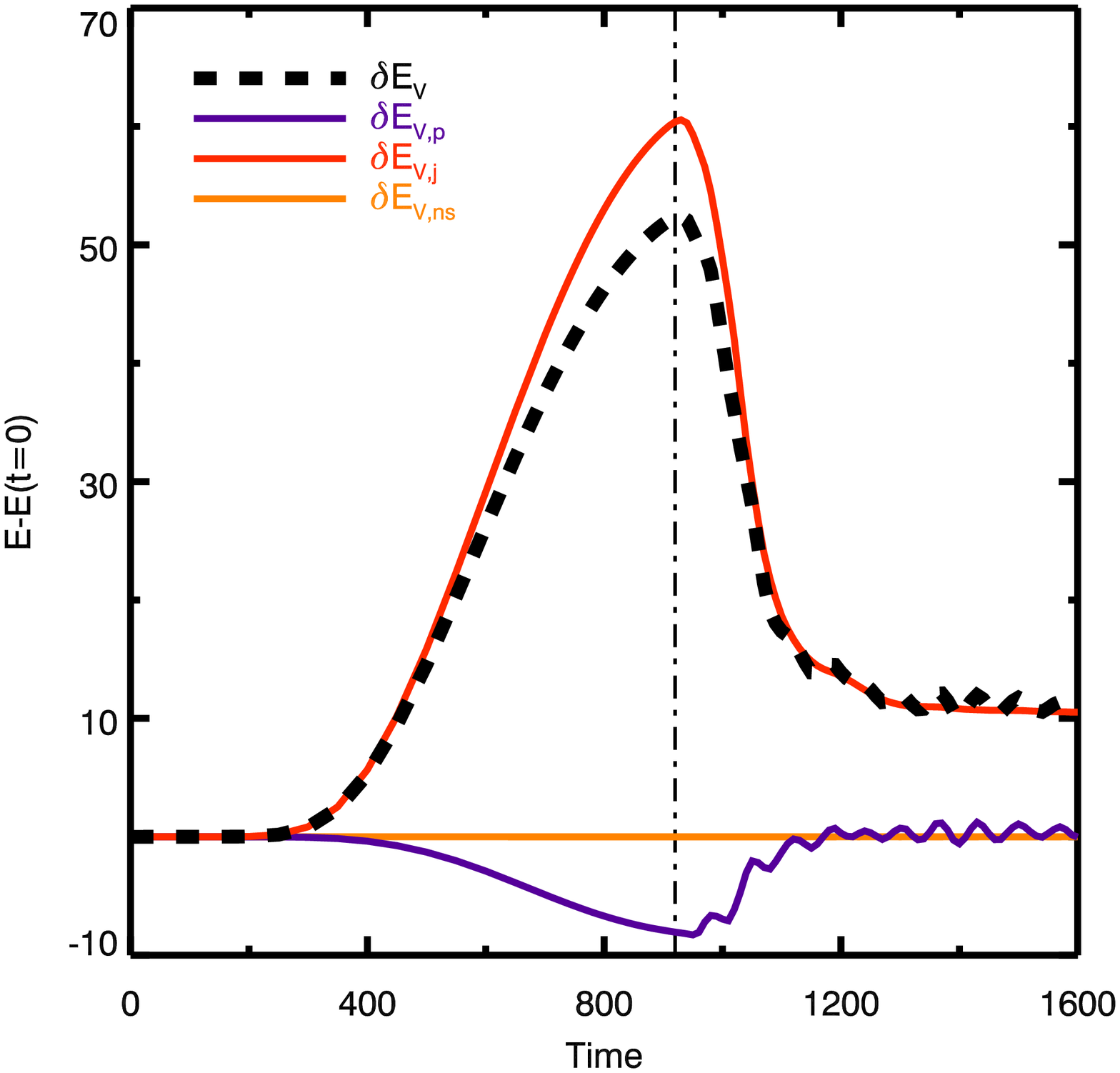}
  \includegraphics[width=\imsize,clip=true]{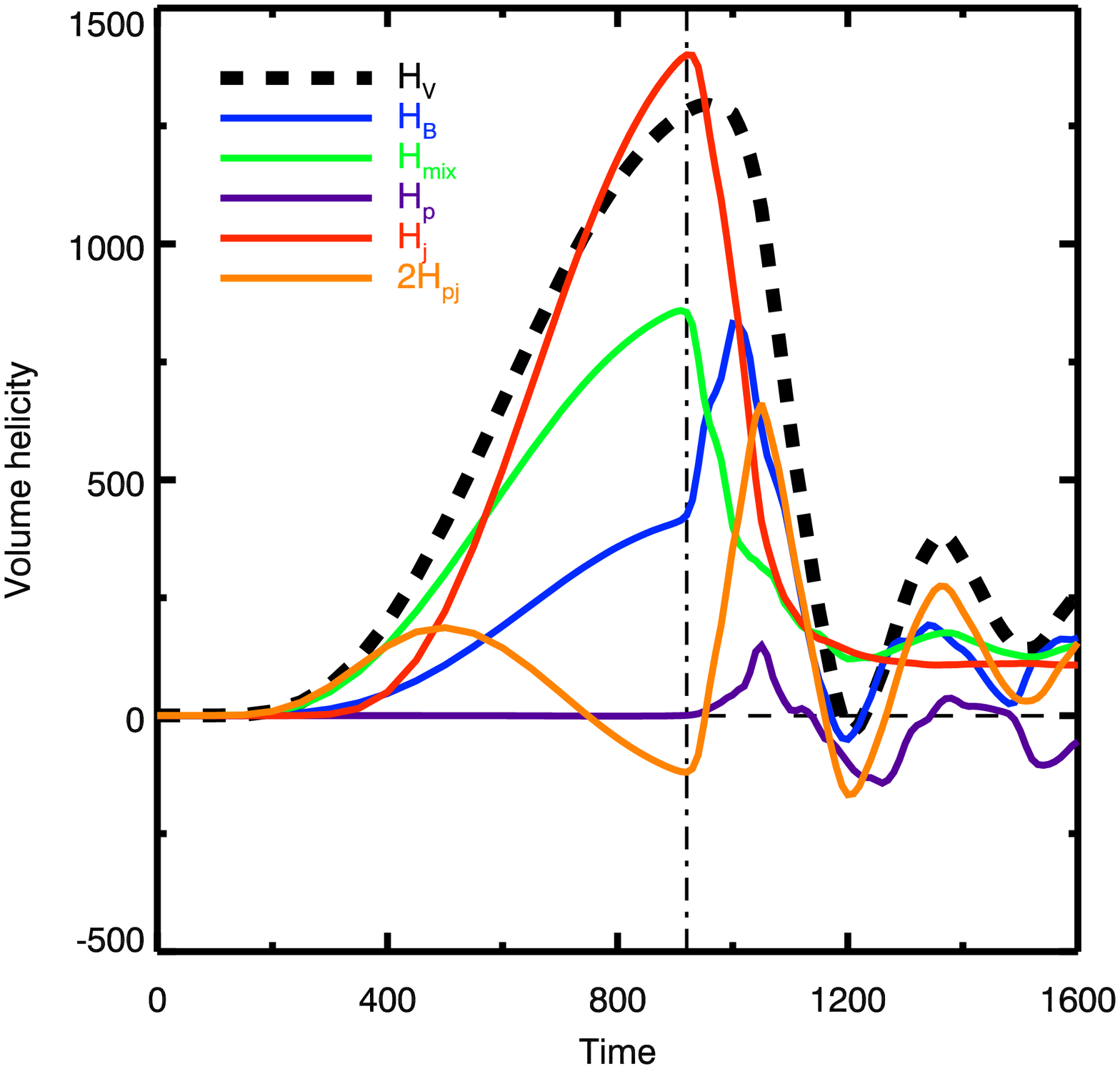}
  \caption{    
    \textbf{Top panel}: Time evolution of the different magnetic energies relative to their respective initial values: total 
    ($\delta E_{\vol}$, black dashed line), of the potential magnetic fields ($\delta E_{\vol,p}$, purple line), of the non solenoidal component 
    ($\delta E_{\vol,ns}$, orange line), and free magnetic energy ($\delta E_{\vol,j}$, red line).
    \textbf{Bottom panel}: Relative magnetic helicity ($H_{\vol}$, black dashed line) and its decompositions, \eqs{HrelDecomp}{HrelDecomp2}, 
    computed with the practical DeVore method: proper helicity ($\Hm$, blue line), potential field helicity ($\Hp$, purple line), and the mixed 
    helicity ($\Hmix$, green line) ; current-carrying helicity ($\Hj$, red line), mutual helicity between potential and current-carrying fields 
    ($\Hpj$, orange line). In both panels, as in all figures hereafter, the dot-dashed vertical line at $t=920$ indicates the transition between the 
    quasi-ideal-MHD/energy accumulation and the non-ideal/jet generation phases.
    }
 \label{Fig:EvolEH}
\end{figure}

\subsection{Volume helicity and flux computations in the DeVore gauge} \label{s:Mhelicomp}

In order to compute magnetic helicity in the volume, $H_\vol$, we will use the method presented in \citet{Valori12}. 
All the vector potentials will be computed using the gauge presented in \citet{DeVore00}, with no vertical component:
\BE 
\label{eq:DeVoreGauge}
A_z = A_{p,z} = 0 
\EE

Under this assumption, the vector potentials can be computed in the volume using a 1D integral \citep[c.f. Equations (10) or (11) of][]{Valori12} with 
a 2D partial differential equation to be solve at the bottom or top boundary \citep[c.f. Equations (9) or (12) of][]{Valori12}.
We tested that we did not obtained significant differences on the helicity conservation and dissipation properties 
whether the integration was done from the top or the bottom boundary of the system. 
In the present manuscript the results were obtained with the integration performed from the top boundary.

While other gauge choices could be explored, choosing the DeVore gauge, \eq{DeVoreGauge}, is here motivated by the fact that it is 
numerically efficient and convenient. This selection leaves a freedom for the gauge of $\vA$ and $\vAp$ which can be independent,
i.e. not linked as in \eq{CondA=Ap}. In our computation of $\vAp$ and $\vA$, we use gauge freedom to additionally 
enforce that the two vectors are equal at the top \citep[see Eq. 29 of][]{Valori12}:
 \BE 
\label{eq:CondAAptop}
\vA (z_{\rm top}) = \vAp(z_{\rm top})
\EE
We still have a freedom on $\vAp(z_{\rm top})$ as expressed by the 2D partial differential equation (20) of \citet{Valori12}.  Here, we select 
their particular solution expressed by their equations (24,25). With this additional choice, both $\vAp$ and $\vA$ are uniquely defined 
(modulo a constant) by vertical integration starting from the top boundary. 

Practically, we first determine the potential field by solving the solution of the Laplace equation~(17) for $\phi$ of \citet{Valori12}. 
Then we can compute the potential vectors from a direct 1D integration of $B_z$ starting from the top boundary 
\citep[c.f. Section 3.3 of][]{Valori12}. We refer to this method below as the "practical DeVore method" since it is efficient and easy to implement. 

While the condition (\ref{eq:CondAAptop}) holds at the top boundary it does not hold at the other boundaries.
Indeed, Eq. 13 in \cite{Valori12} can be used to show that the difference between the value of $\vAp$ and $\vA$ at the bottom boundary is equal to 
$\hat{\vec{z}}\times\int_{z_1}^{z_2}(\vB-\vBp)\dV$. It is important to note that the 
conditions of $\Hmix=0$ and  $\vAp \cdot  \dS|_\surf=0$ are never enforced in this gauge. The relative helicity terms $\Hmix$ and $\Hp$ and the 
helicity flux term $\FAAp$ and $\Fphi$ can thus never be considered null \citep[see also Section 3.4 in][ for more details]{Valori12}.
Therefore, for the practical DeVore method, the helicity variation of the system is given by the general formula of \eq{Hvar4}.

Alternatively, we will determine the helicity assuming a Coulomb gauge for $\vAp$ only. The DeVore and Coulomb gauge are indeed compatible 
for a potential field. We can solve the 2D partial differential equation (20) of \citet{Valori12} as a Poisson problem
to obtain $\vAp$ (see their Eq.~(41), but translated to the top boundary). This implies by construction that $\vAp$ is simultaneously
respecting the DeVore, \eq{DeVoreGauge}, and the Coulomb gauge, \eq{Coulomb}. In the following, we will refer to this method as the 
``DeVore-Coulomb method''. 

While still following the DeVore condition (\ref{eq:DeVoreGauge}), $\vA$ computed in 
the DeVore-Coulomb method will be different from $\vA$ computed in the practical DeVore method
because the boundary condition, \eq{CondAAptop} is different. In particular, the distributions of $\vA$ at the bottom boundary will 
be significantly different with each methods, hence leading to very different values of $\FBn$ and $\FVn$.
In Appendix \ref{s:GenDev-Coul}, we present an additional test, with $\vAp$ computed in the Coulomb gauge but where $\vA$ is not satisfying 
the condition (\ref{eq:CondAAptop}).

\subsection{Test data set}\label{s:Msimulations}

In order to test the helicity conservation, we employ a test-case 3D MHD numerical simulation of the generation 
of a solar coronal jet \citet{Pariat09a}.
Figure~\ref{Fig:jetsim} presents snapshots of the evolution of the magnetic field. 
The simulation assumes an initial uniform coronal plasma with an axisymmetric null point magnetic configuration (left panel).
The magnetic null point is created by embedding a vertical dipole below the simulation domain, and adding an uniform volume vertical 
magnetic field of opposite direction in the domain. The null point present a fan/spine topology, dividing the volume in two domains of 
connectivity, one closed surrounding the central magnetic polarity, and one open. 

The computation are performed with non-dimensional units. We analyze the time evolution of the magnetic field from $t=0$ to $t=1600$. The time
steps between two outputs is $\Delta t=50$ for $t\in[0;700]$, during the accumulation phase, and, $\Delta t=10$ for $t\in[700;1600]$ 
during the dynamic phase of the jet. 
The analysis volume $\vol$ is a subdomain of the larger discretized volume employed in the original MHD simulation:  $[-6,6]$ in $x$-, $[-6,6]$ 
in $y$-, and $[0,12]$ in the $z$-directions, thus only taking into account the region of higher resolution \citep[c.f. Figure 1 of][]{Pariat09a}.

The ideal MHD equations are solved with the ARMS code based on Flux Corrected Transport algorithms \citep[FCT,][]{DeVore91}. 
The parallelisation of the code is ensured with the PARAMESH toolkit \citep{MacNeice00}. In the original simulation of \citet{Pariat09a}, 
reconnection is strongly localized thanks to the use of adaptive mesh refinement methods at the location of the formation of the thin current sheets involved in reconnection \citep[c.f. Appendix of][]{Karpen12}. 
In the present paper in order to keep the resolution of the domain constant we have however switched off adaptivity, so the resolution is equal to the 
initial one \citep[as in Figure 1 of][]{Pariat09a}, throughout the whole simulation. 

The variation of energies relatively to their initial value, $\delta \E(t)=\E(t)-\E(t=0)$ is displayed in \fig{EvolEH}, top panel. The index $\vol$ indicates that 
the energy is computed by a volume integration. The total energy, $\E$, in the domain can be decomposed as:  
\BE 
\label{eq:thomson}
E = \Ep +\Ej +E_{\vol, \rm ns} \,,
\EE
where $\Ep$ and $\Ej$  are the energies associated to the potential and current-carrying solenoidal contributions and 
$E_{\vol,ns}$ is the sum of the nonsolenoidal contributions \citep[see Eqs.~(7,8) in][ for the corresponding expressions]{Valori13}.
Initially the system is fully potential and $E_\vol=E_{\vol,p}$. Energy is injected in the system by line-tied twisting motions of the central polarity. The 
axisymmetric boundary motions are preserving the distribution of $B_z$ at the bottom boundary.
Magnetic free energy and helicity accumulates monotonically increasing the twist in the closed domain (\fig{jetsim}, central left panel).
The potential field energy $E_{\vol,p}$ decreases slightly during the accumulation phase because of the bulge of the central domain which is changing 
the distribution of the field on the side and top boundaries. The $E_{\vol,ns}$ term is almost constantly null, an indication of the excellent solenoidality 
of the system 

Eventually, around $t \simeq 920$, the system becomes violently unstable: magnetic reconnection sets in and the closed twisted field lines 
reconnect with the outer open field lines. A steep decrease of the free magnetic energy is observed in \fig{EvolEH}: 83\% of the maximum free 
magnetic energy is dissipated/ejected/transformed. 
Through reconnection, twist and helicity are expelled from the central domain, inducing a large scale kink wave which exits through the top boundary 
(\fig{jetsim}, central right panel). This large scale non-linear magnetic wave, simultaneously compresses and heats the plasma, inducing the generation 
of an untwisting jet that can observationally be interpreted as a blowout jet \citep{Patsourakos08,Pariat15}. 
The driving motions had been slowly ramped down at the 
time of the trigger of the jet so that few energy and helicity are injected from the lower boundary after the jet onset \citep[cf. Figure 6 of][]{Pariat09a}. In 
the final stage the system slowly relaxes to a configuration similar to its initial state, with the potential field energy being similar to its initial value, and 
only few field lines remaining twisted, next to the inversion line (\fig{jetsim}, right panel).  

This simulation thus presents two distinct phases typical of active events: before $t \simeq 920$, a phase with a slow ideal accumulation of 
magnetic helicity and energy, and after $t \simeq 920$, an eruptive phase of fast energy dissipation and helicity transfer involving non-ideal effects. 
In the first phase, the system behaves very close to ideality as 
demonstrated in a benchmark with a strictly ideal simulation \citep{Rachmeler10}. In the non-ideal phase, \cite{Pariat09a} showed that 90\% of the 
helicity was eventually ejected through the top boundary by the jet, and \cite{Pariat10} showed the high reconnection rate processing the 
magnetic flux during the jet. These two phases allow us to test the conservation of helicity in two very distinct paradigms of MHD.

In the following we test the helicity conservation with the above MHD simulation. We first use the practical DeVore method (see \sect{DeVore}). 
We next  test the effect of the gauge choice on our results, by using the DeVore-Coulomb method (see \sect{DeVore-Coulomb}). 
We will show how the gauge choice possibly affects the evolution of each terms. 

\begin{figure}[ht]
  \setlength{\imsize}{\textwidth}
  \includegraphics[width=0.5\imsize,clip=true]{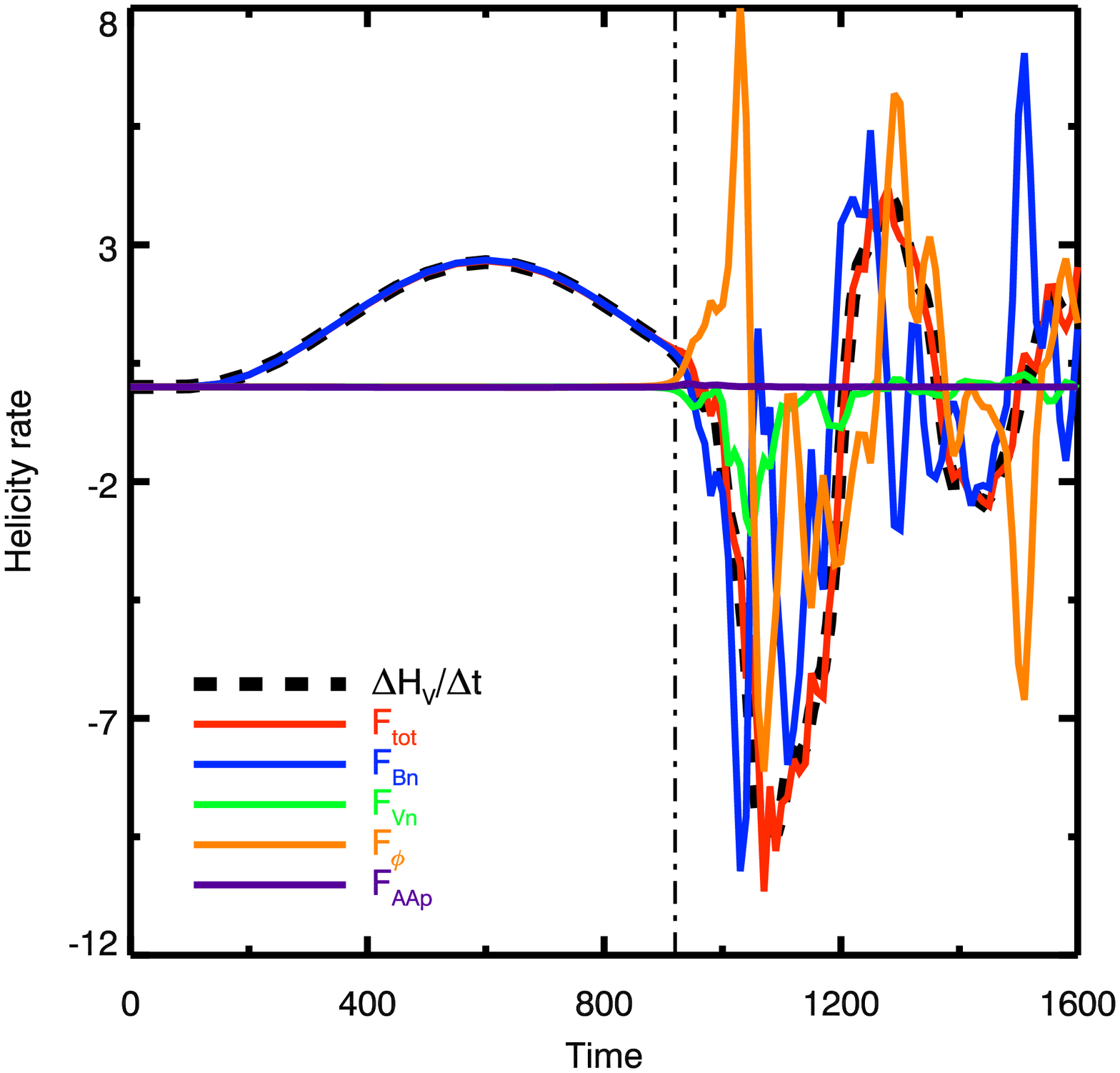}
  \caption{ Comparison of the helicity variation rate and the helicity flux computed in the practical DeVore gauge. 
  The helicity variation rate ($\Delta H_\vol/\Delta t$, dashed line) is derived from the volume integration method. The total helicity fluxes 
  through the whole surface, $\surf$, of the domain is $\Ftot$ (red line). This flux is composed of, \eqss{defFVn}{defFphi}:
  $\FVn$ (green line),  $\FBn$ (blue line), $\Fphi$ (orange line), $\FAAp$ (purple line). 
                      }
  \label{Fig:EvolHflux}
\end{figure}

\begin{figure*}[ht]
  \setlength{\imsize}{0.33\textwidth}
 \includegraphics[width=\imsize,clip=true]{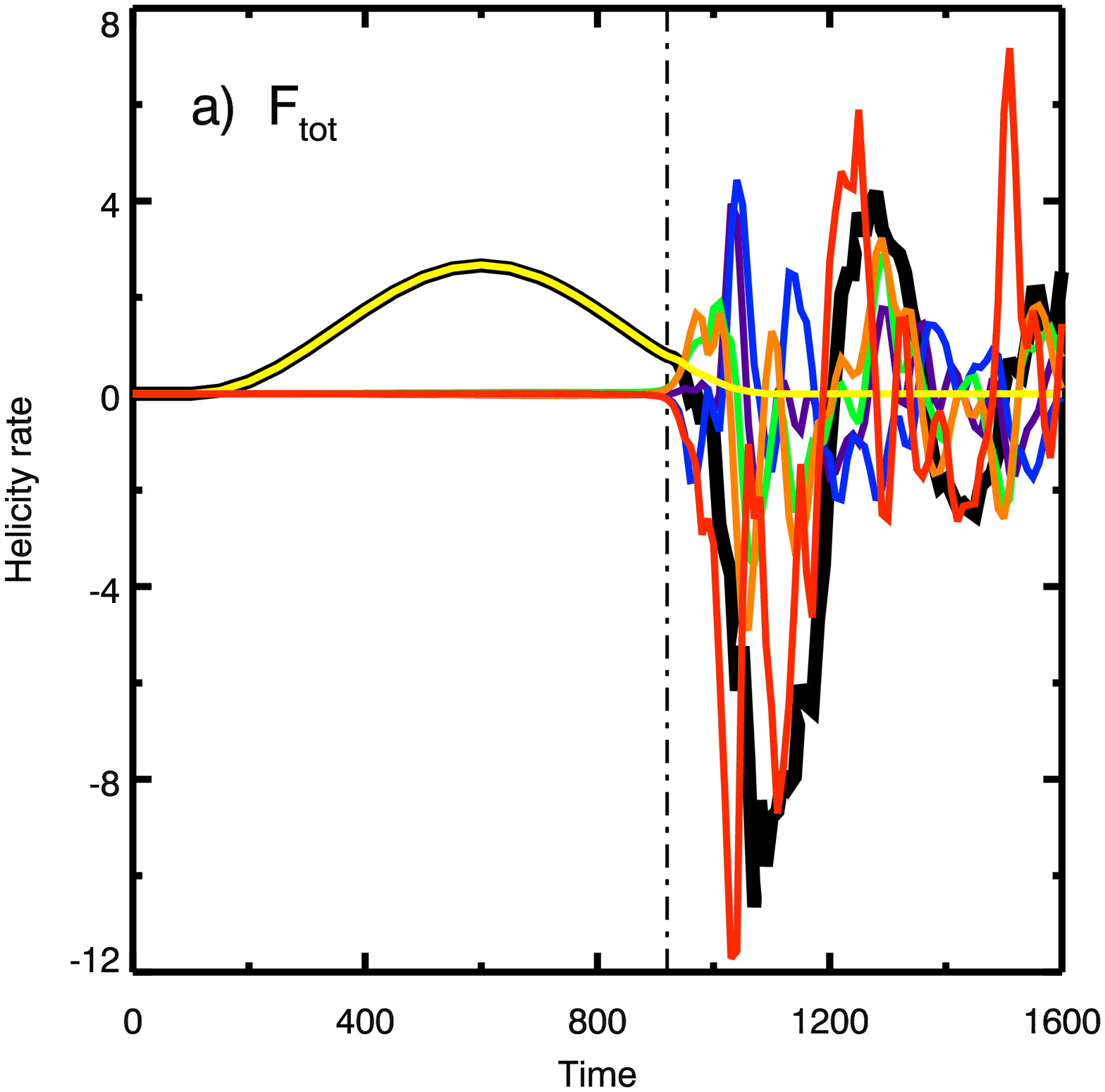}
  \includegraphics[width=\imsize,clip=true]{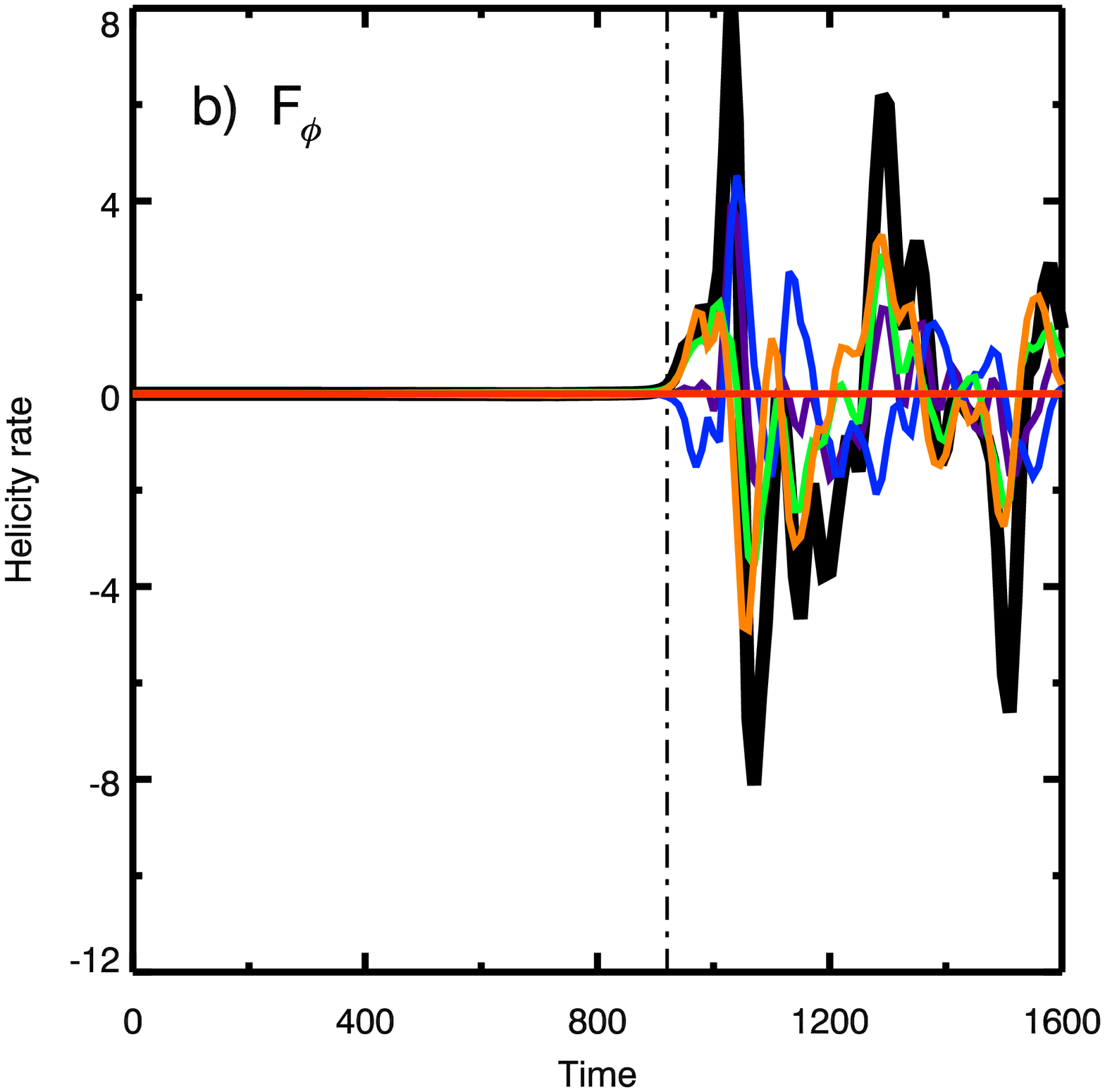}
 \includegraphics[width=\imsize,clip=true]{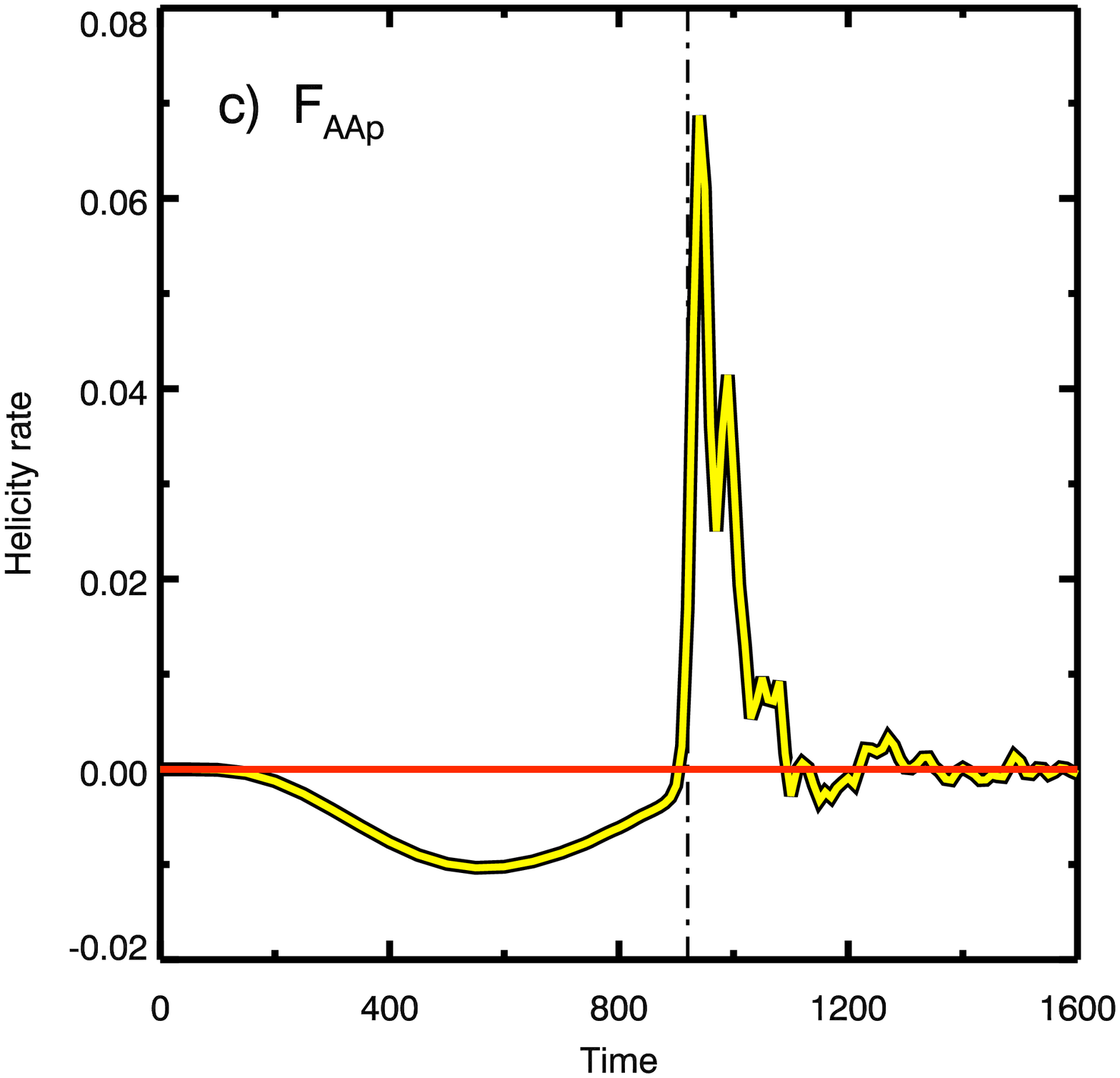}
\includegraphics[width=\imsize,clip=true]{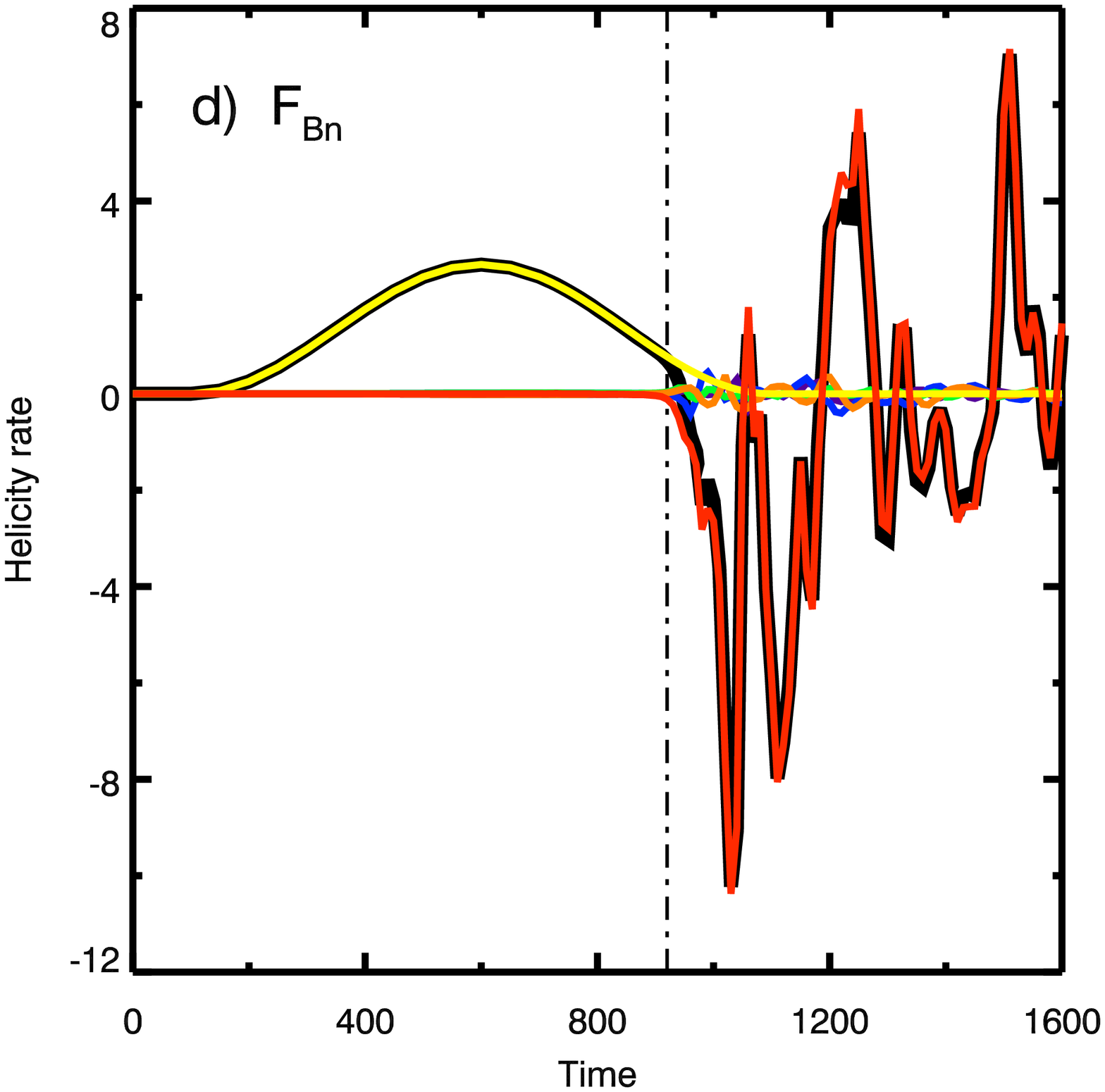}
  \includegraphics[width=\imsize,clip=true]{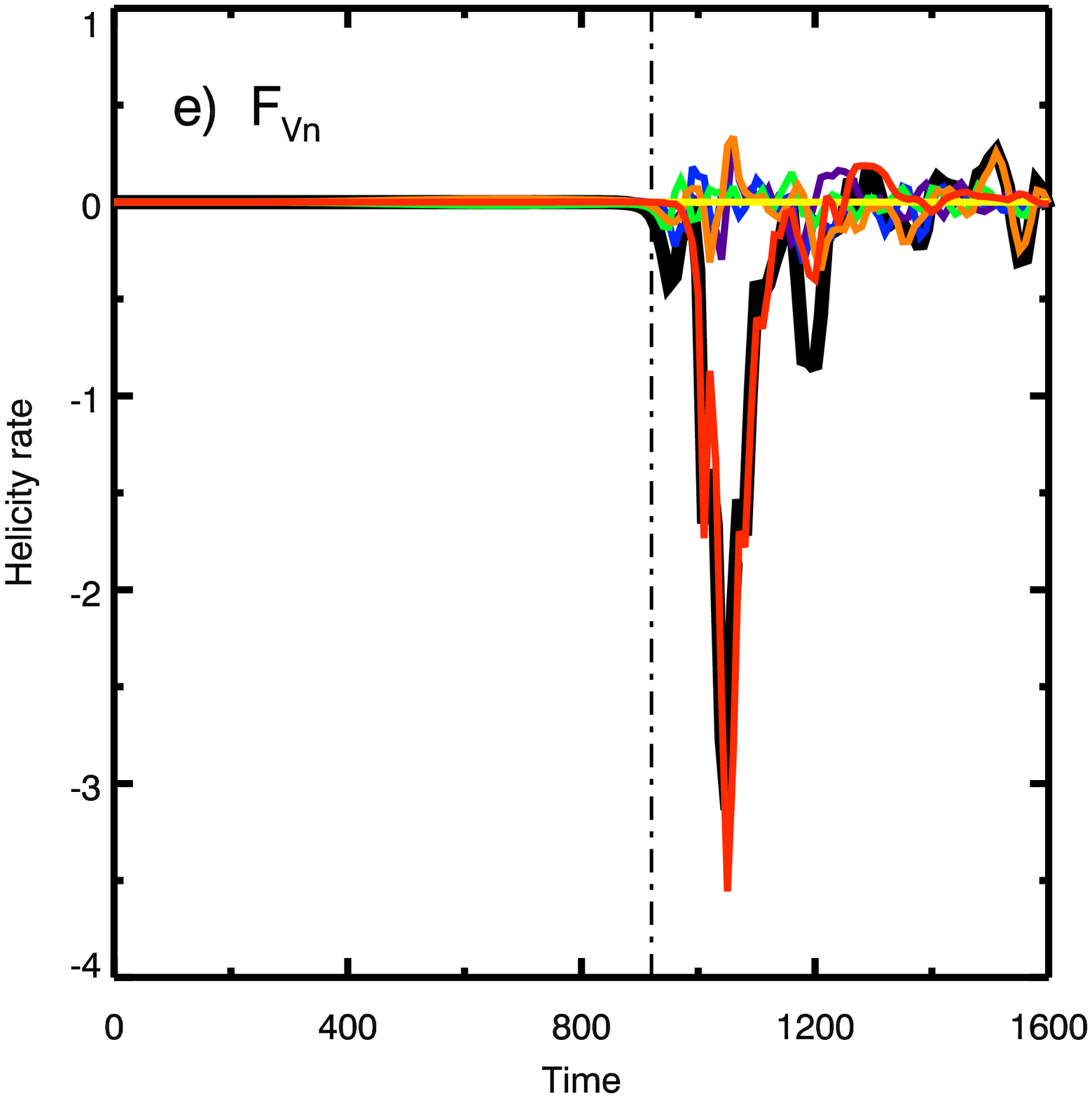}
  \includegraphics[width=\imsize,clip=true]{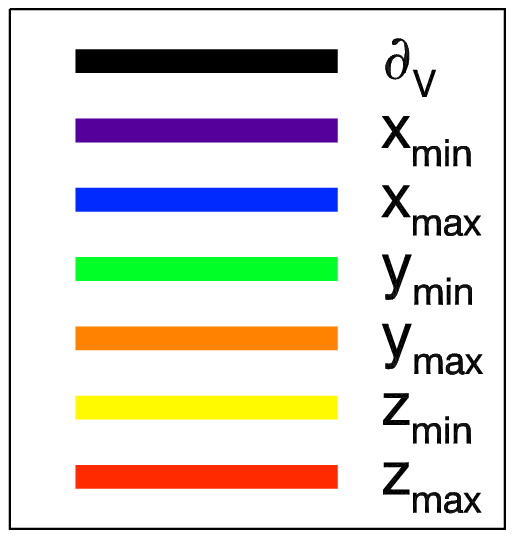}
 \caption{Total helicity flux, $\Ftot$ (Eq.~\ref{eq:defFtot}), and the the terms compositing it, \eqss{defFVn}{defFphi}, through the different boundaries 
 computed in the practical DeVore Gauge.
 \textbf{(a)}:  $\Ftot$;  \textbf{(b)}:  $\Fphi$; \textbf{(c)}:  $\FAAp$; \textbf{(d)}: $\FBn$;  \textbf{(e)}: $\FVn$. 
 In each plot the dark line correspond to the sum of the flux through all the boundaries while the color lines correspond to a flux through a 
 particular boundary (purple and blue: left and right $x$ sides, green and orange: front and back $y$ sides; yellow and red: bottom and top $z$ sides).
}
  \label{Fig:Hfluxbnd}
\end{figure*}

\section{Magnetic helicity conservation in the practical DeVore gauge} \label{s:DeVore}

Our first study of the helicity variations is performed in the practical DeVore gauge. The only assumptions on the potential field and the vector potentials 
are given by Equations (\ref{eq:condrelhgauge}),  (\ref{eq:DeVoreGauge}) \& (\ref{eq:CondAAptop}). 
This corresponds to a general case where the helicity variation of the system is provided by \eq{Hvar4}.

\subsection{Helicity evolution}\label{s:DevHevolution}

Figure \ref{Fig:EvolEH}, bottom panel, presents the evolution of the relative magnetic helicity $H_\vol$ in the system. Similarly to magnetic energy the 
two phases of the evolution are clearly marked. The first phase corresponds to a steady accumulation of magnetic helicity, while the second 
corresponds to the blowout jet. 
The latter is associated with a steep decrease of magnetic helicity. As the system relaxes the magnetic helicity value oscillates. These 
oscillations are related to the presence of a large scale Alfv\'enic wave which is slowly damped after the jet. This oscillations can also be seen in the total energy but with a much smaller relative amplitude (so small that $\E$ still decreases 
monotonically). 

The right panel of \fig{EvolEH} also presents the decomposition of the relative magnetic helicity, $H_\vol$, in $\Hm$, $\Hmix$ and $\Hp$ of \eqss{HrelDecomp}{defHmix} .  During the 
accumulation phase, $H_\vol$ is dominated by $\Hmix$ while $\Hp$ is almost constantly null. During the jet, one can see strong fluctuations of the relative importance of these terms. $\Hm$, $\Hmix$ and $\Hp$ eventually present contributions of similar 
amplitude. However, because these terms are not gauge invariant, their value in a different gauge might be quite different (cf. \sect{DeVore-Coulomb}).

The decomposition of $H_\vol$ with $\Hj$ and $\Hpj$ of \eqss{HrelDecomp2}{defHpj} is also plotted in \fig{EvolEH}, bottom panel. $\Hj$ has an evolution comparable to $\delta \Ej$. It captures most of the helicity evolution both during the accumulation and jet phases. In contrast, the mutual helicity between the potential and the current carrying fields contains mostly oscillations. Therefore, $\Hj$, which is a gauge invariant quantity, is a promising quantity to analyze during a jet/eruption. 

\begin{figure*}[ht]
  \setlength{\imsize}{0.35\textwidth}
  \sidecaption    
  \includegraphics[width=\imsize,clip=true]{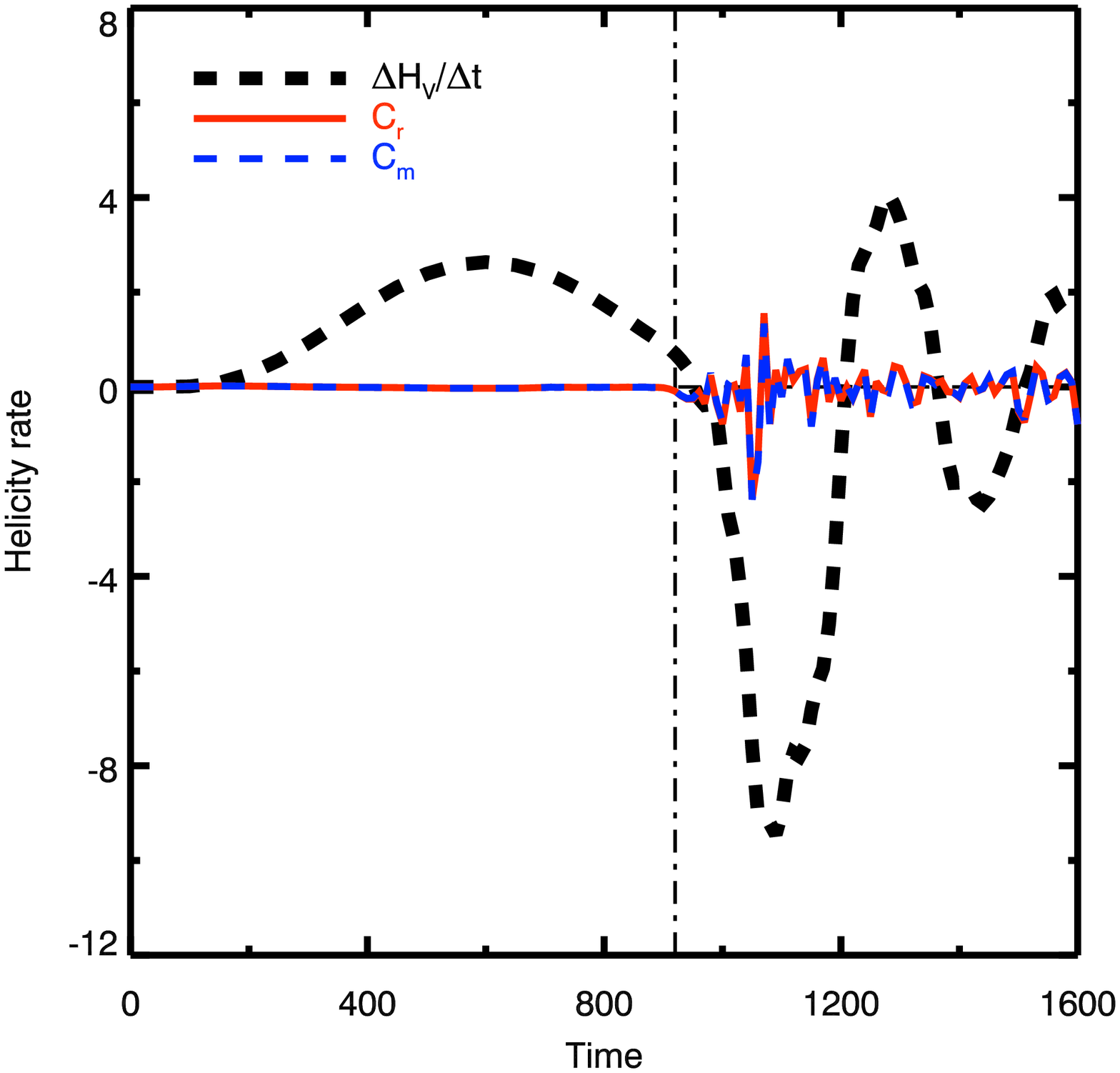}
  \includegraphics[width=\imsize,clip=true]{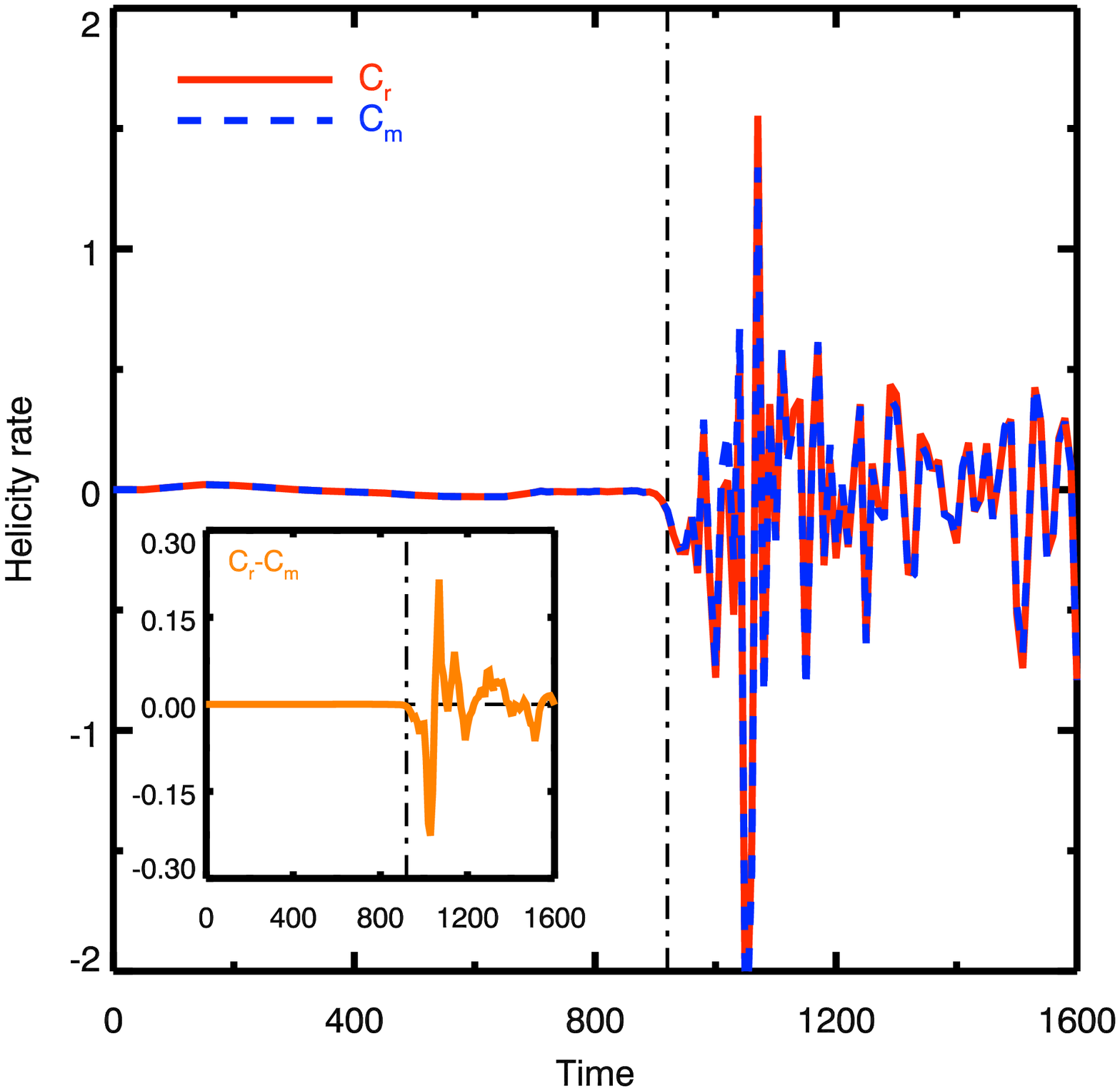}
  \caption{
  Difference between the helicity variation rate and the boundary helicity flux, computed in the 
  practical DeVore gauge.  The relative helicity conservation criterion $\Cr$  (red line, Eq.~\ref{eq:defCr}) and the magnetic helicity dissipation criterion $\Cm$ (blue line, Eq.~\ref{eq:defCm}) are plotted relatively to the helicity variation rate ($\Delta H_\vol/\Delta t$, black dashed line, left panel) and in their own amplitude range (right panel). The insert 
in the right panel present the difference between $\Cr$ and $\Cm$, i.e. the potential helicity volume variations $\dHdt_{\rm p,var}$.
}
  \label{Fig:Hvolfluxdiff}
\end{figure*}

\subsection{Helicity fluxes}\label{s:DevHfluxes}

We compute the time variations $\Delta H_\vol/\Delta t$ of the helicity determined with the volume integration method, and compare it with the 
different terms of the relative magnetic helicity flux through the whole system boundary (\fig{EvolHflux}). This shows that the helicity variation 
$\Delta H_\vol/\Delta t$ is very closely matching the curve of $\Ftot$, indicating that the variation of helicity in the domain are tightly related to 
the flux of relative helicity through the boundary. The core results of this study is that indeed magnetic helicity is very well conserved in the 
studied simulation, both during the quasi-ideal-MHD and non-ideal phases.

During the non-ideal/jet phase, strong fluctuations are observed for all terms but the $\FAAp$ term. The later is constantly negligible 
relatively to the others. On the other hand we see that the $\Fphi$ term displays important fluctuations, frequently of similar amplitude 
and opposite sign to $\FBn$.  The $\Fphi$ term clearly cannot be neglected in this gauge. 

The analysis of the flux of each terms through each individual boundary is insightful (\fig{Hfluxbnd}), though one must bear in mind that the 
plotted terms are not gauge invariants.  Although its amplitude is extremely small, a finite flux of $\FAAp$ is present only at the bottom 
boundary during the whole evolution of the system: there is no flux on the sides because of the DeVore gauge 
(Eq.~\ref{eq:DeVoreGauge}), and no flux on the top because of the imposed condition of \eq{CondAAptop}.

During the ideal phase, the flux of helicity is completely dominated by the $\FBn$ term originating from the bottom photospheric boundary. This is 
consistent with the fact that the system is indeed driven by horizontal shearing motions at the bottom boundary. No remarkable helicity flux is observed 
at the other boundaries during this period. Helicity is thus accumulating in the volume $\vol$. 

During and after the jet ($t > 920$) important helicity fluxes are noted in the side and top boundaries, while the bottom flux is now negligible 
as the boundary flows have been ramped down in amplitude. A large flux of helicity occurs at the top boundary (red curves), dominated 
by the $\FBn$ term and to a lower extend by the $\FVn$ term. This corresponds to the ejection of helicity by the jet, driven by a 
large-scale non-linear torsional wave \citep{Pariat09a,Pariat15}. 
$\FVn$ is peaked at the time of the passage of the bulk of the jet through the top boundary.  
The flux of $\FBn$ and $\FVn$ through the side boundaries, while present, is comparatively small. However, the side boundaries see the 
transit of important flux $\Fphi$. No specific side boundary is dominating the total value of $\Fphi$.
Due to the DeVore gauge (Eq. \ref{eq:DeVoreGauge}), $\Fphi$ is null at the bottom and top boundaries.  

We conclude that, computed with the particular DeVore method, the total flux of helicity $\Ftot$ during the jet consists of complex transfer of helicity through all the boundaries of the system.

\begin{figure*}[ht]
  \setlength{\imsize}{0.35\textwidth}
  \sidecaption    
  \includegraphics[width=\imsize,clip=true]{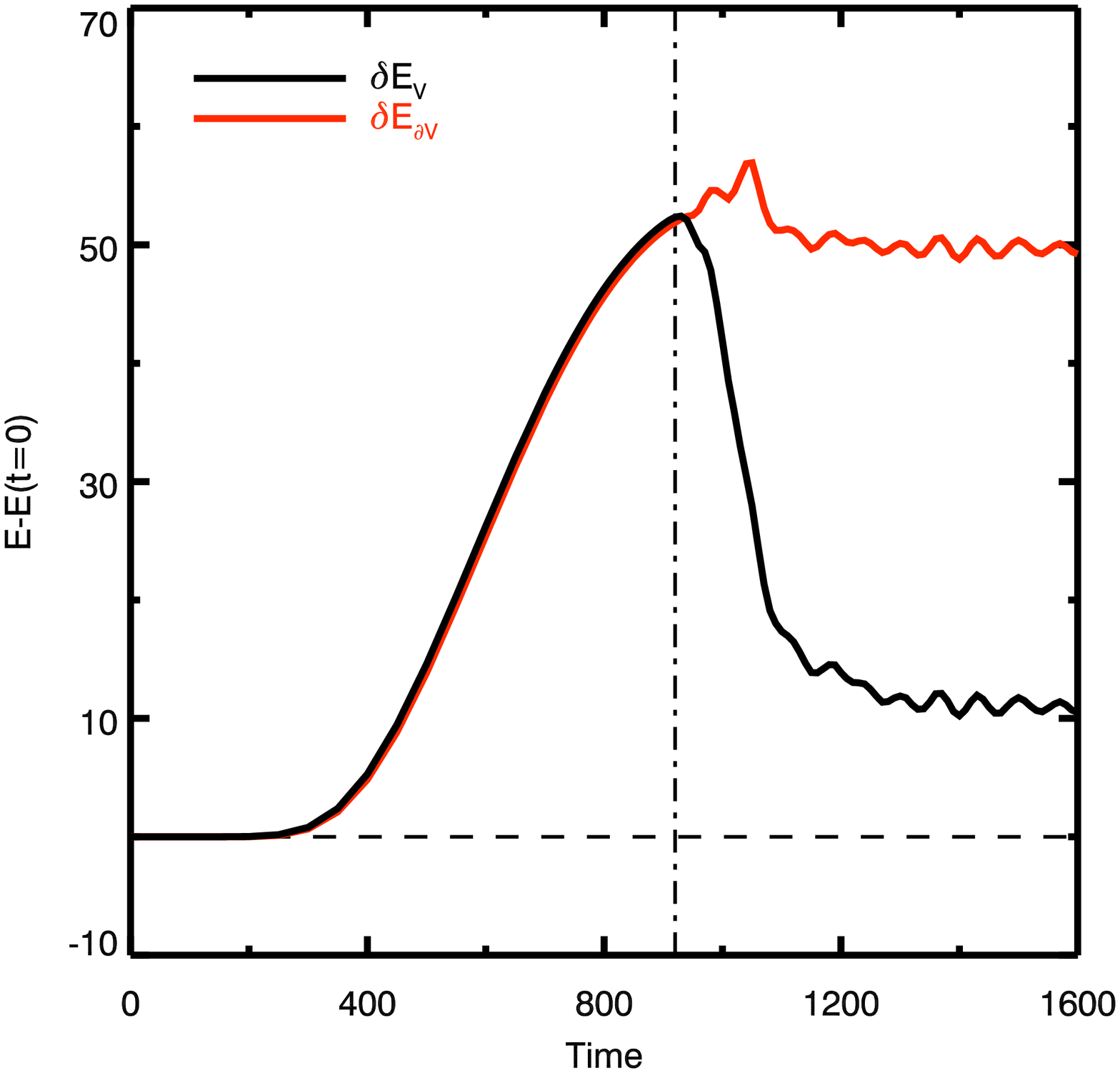}
  \includegraphics[width=\imsize,clip=true]{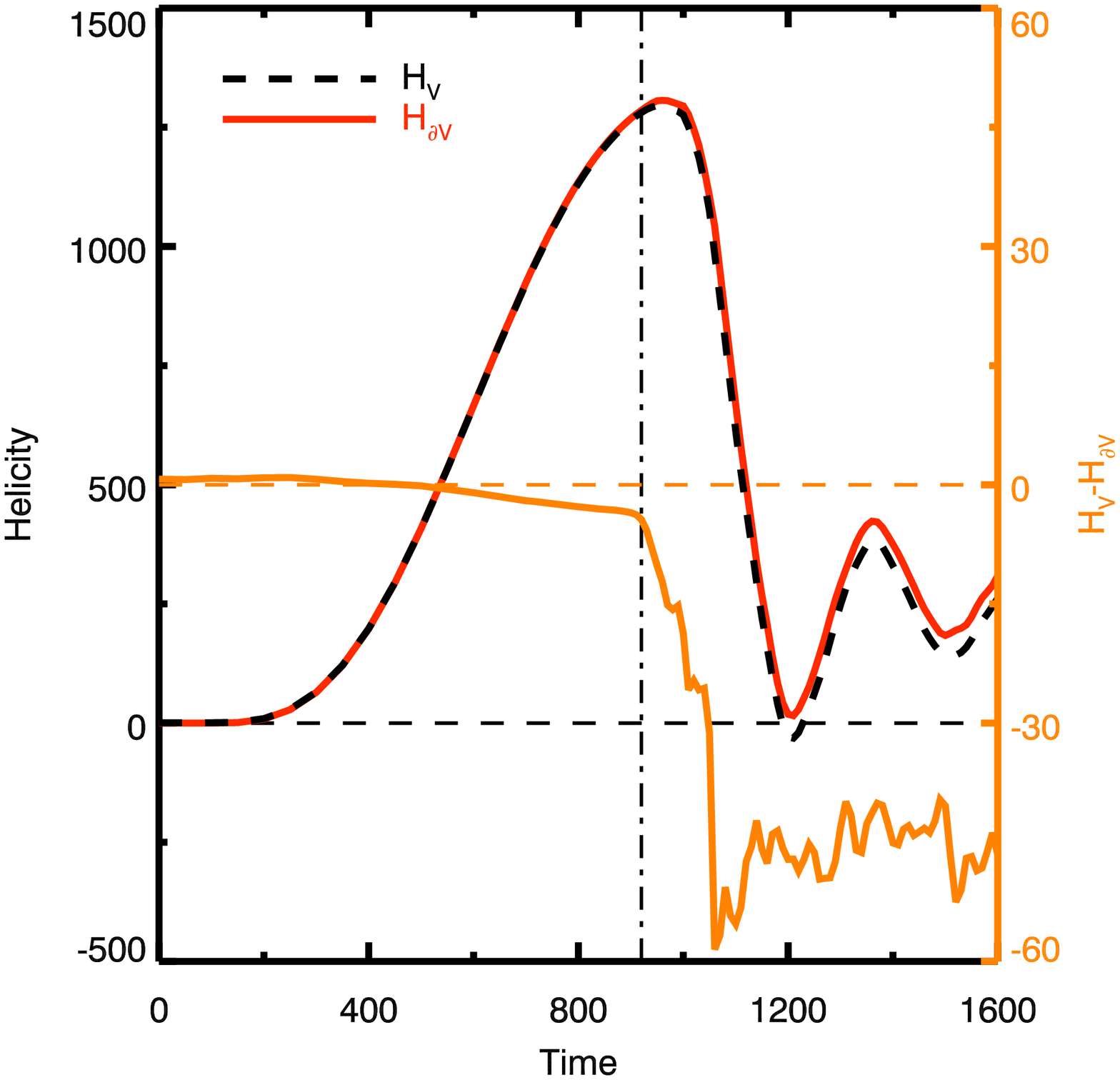}
  \caption{\textbf{Left panel}:  Comparison of the evolution of magnetic energy ($\delta E_\vol$, black line) 
in the volume with the time integration of 
  the Poynting flux  through the whole surface of the domain ($\delta E_\surf$, red line). \textbf{Right panel}:
  Comparison of the evolution of helicity obtained by volume integration ($H_\vol$, black dashed line) with the time integration of the helicity flux 
  through the whole surface of the domain ($H_\surf$, red line), computed with the practical DeVore gauge. Their difference is plotted in orange 
  on a different range of amplitude (\cf\ right axis).
                      }
  \label{Fig:EHdiss}
\end{figure*}

\subsection{Helicity conservation}\label{s:DevHconservation}

In order to better estimate the helicity conservation and dissipation, we have plotted in \fig{Hvolfluxdiff} the criterions $\Cr$ and $\Cm$ 
(see their definition in \sect{Mestimators}). We observe that these two criterions are almost equal. They differ only by the term describing 
the volume variation of the potential helicity $\dHdt |_{\rm Bp,var}$ of \eq{defFVn}. Our calculation finds that $\dHdt |_{\rm Bp,var}$ is 
extremely small compared to the variation of $\Cr$ and $\Cm$ (see inset in the right panel of \fig{Hvolfluxdiff}), even though we are 
not explicitly enforcing $\divAp =0$. This term is smaller than the combined effect of the real helicity 
dissipation $\dHdt |_{\rm diss}$ and the numerical errors on the volume and flux helicity measurements.

The curve of $\Cr$ demonstrates that magnetic helicity is extremely well conserved during the ideal-MHD phase of the simulation and that 
it is also very well conserved during the non-ideal phase. Moreover, during the ideal phase the amplitude of 
$|\Cm|$ does not exceed $0.029$ which is $1\%$ of  the maximum amplitude of helicity variation during the period. 
At the end of this period, one also have $\epscm (t=920) <1\%$, thus helicity is very weakly dissipated, as theoretically expected \citep{Woltjer58}. 

During the non-ideal phase, $\Cr$ and $\Cm$ show high frequency fluctuations around the null value, decorrelated from the fluctuation of helicity in 
the system. Our analysis indicates that while these fluctuations could originate from the real physical term $\dHdt |_{\rm diss}$, 
they are in fact dominated by the numerical precision on the estimation of $\Ftot$. From the same simulations data, we will see 
in \sect{DeVore-Coulomb} that the computation with the DeVore-Coulomb method reduces these fluctuations.

In \fig{EHdiss} we have plotted the variation of magnetic energy and helicity in the system computed with a volume integration and from the integration 
of the Poynting and helicity fluxes through the boundaries of the system. During the ideal-MHD phase both magnetic energy and helicity are well 
conserved, their volume variations being equal to their boundary fluxes. During the non-ideal phase, while magnetic helicity is still very well conserved 
magnetic energy is clearly not.  When the jet is generated, the magnetic energy quickly decreases: part of it is ejected through the top boundary by the 
jet, but for most part is dissipated in the reconnection current sheet and transformed in other forms of energy.  When the simulation is 
stopped, we determined that about $17\%$ of the magnetic energy injected in the system by the bottom boundary motions remains in the system,  
$21\%$ is directly ejected with the jet and the $62\%$ remaining are dissipated/transformed in other form of energy \citep[see also][]{Pariat09a} 
As expected during reconnection, magnetic energy is strongly non-conserved. 

The situation for magnetic helicity is very different. At the end of the ideal phase the maximum difference between 
$H_\vol$ and $H_\surf$ is of $3.5$ units. 
During this period a total of $H_{ref}=max(H)=1265$ units have been injected and only a fraction $\epsH(t=920) =0.3\%$ is lost. 
The ideality of the system is thus very well maintained. 
During the non-ideal phase, the maximum difference between $H_\vol$ and $H_\surf$ is equal to $58$ units (orange line, \fig{EHdiss}, right panel) and 
$\epsH (t=1600)=4.5\%$. 
The jet is able to transport away a huge fraction, $77\%$ of the helicity of the system. Finally, about $19\%$ of the helicity remains in the 
system at $t=1600$ while $H_\vol$ is still oscillating slightly (\fig{EHdiss}, right panel).

Summarizing, with a generic gauge, for a solar-like active events, we have thus confirmed \citet{Taylor74} hypothesis that magnetic helicity is very 
well conserved even when non-ideal processes are acting. The relative helicity dissipation is 15 times smaller that the relative magnetic 
energy dissipation. However, with the practical DeVore Gauge, we observed that $\Cm$ is strongly 
fluctuating, possibly limited by numerical precision. We now want to test whether we can obtain better results with a different gauge, 
simultaneously allowing us to test the gauge-invariance of the different helicity variation terms.

\section{Magnetic helicity conservation in the DeVore-Coulomb case} \label{s:DeVore-Coulomb}

Relative magnetic helicity has been defined as a gauge-invariant quantity (see \sect{theoryHrelvar}). However, the surface flux of relative helicity is not, 
neither are the individual terms that defines the flux. In the following we will study the influence of computing magnetic helicity using a different gauge.
The vector potential of the potential field, $\vAp$, is now computed in the DeVore-Coulomb gauge, but not $\vA$ (since the Coulomb gauge is 
only compatible with DeVore gauge for a potential field). Because we also impose the condition of \eq{CondAAptop}, $\vA$ is however also 
recomputed during our DeVore-Coulomb method (see \sect{Mhelicomp}). The case where \eq{CondAAptop} is not enforced is briefly discussed 
in \sect{GenDev-Coul}

\begin{figure*}[ht]
  \setlength{\imsize}{0.35\textwidth}
  \sidecaption    
  \includegraphics[width=\imsize,clip=true]{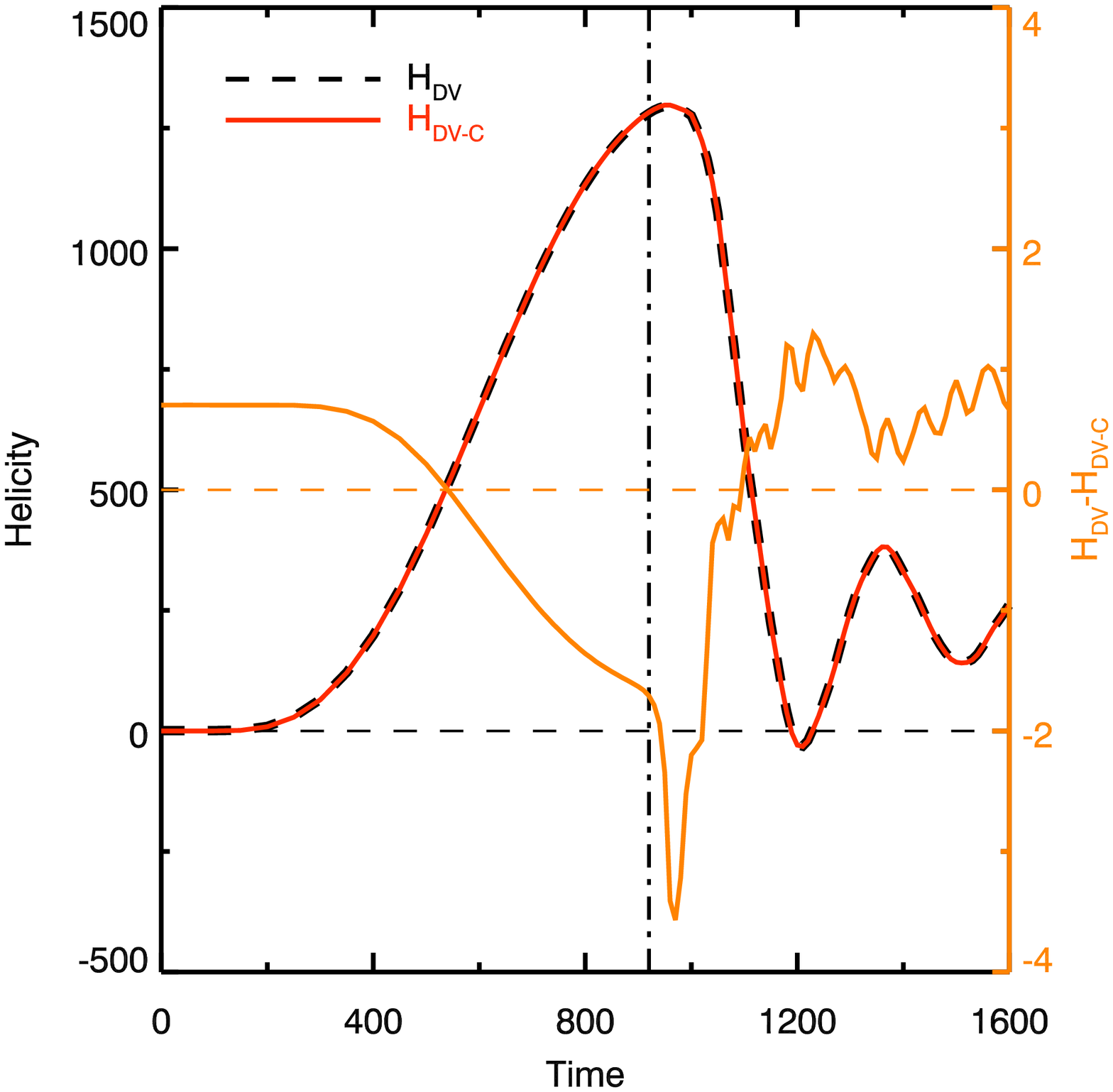}
  \includegraphics[width=\imsize,clip=true]{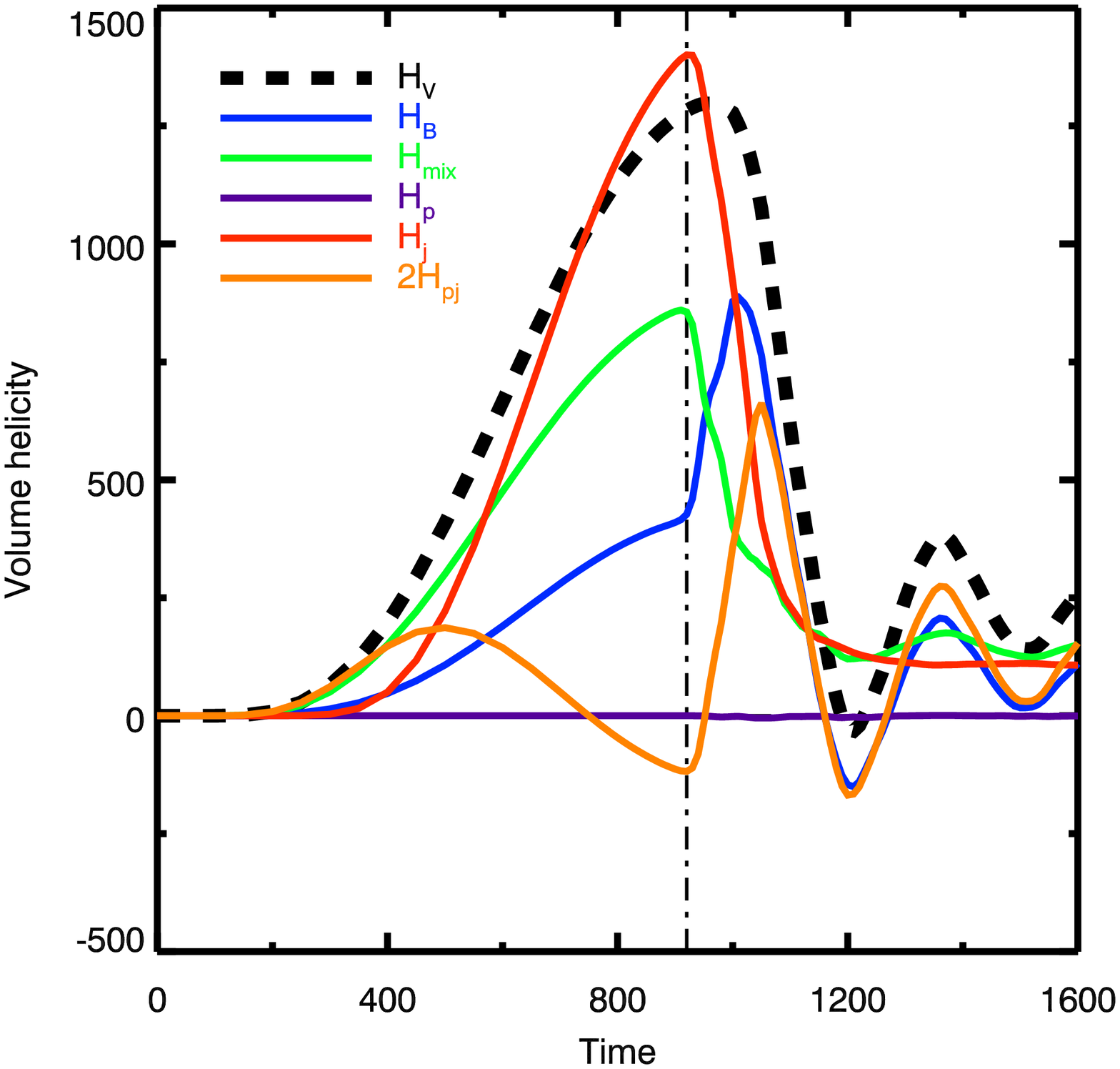}
  \caption{    
     \textbf{Left panel}: Evolution of the magnetic helicity computed with the practical DeVore method (dashed line) and with the DeVore-Coulomb method (red line). Their difference is plotted with an orange line (\cf\ right axis).
     \textbf{Right panel}: Volume magnetic helicity ($H_{\vol}$, black dashed line) and its decomposition in the DeVore-Coulomb case. The plotted curves are similar to Figure \ref{Fig:EvolEH}, bottom panel. 
}
  \label{Fig:GaugecompEvolH}
\end{figure*}

\subsection{Gauge dependance}\label{s:Dev-CoulHevolution}

In \fig{GaugecompEvolH}, we have plotted the evolution of the relative magnetic helicity, $H_\vol$ in the DeVore-Coulomb gauge for $\vAp$.
The left panel shows the comparison with the same quantity computed using the practical DeVore gauge, used in the previous section. We observe that the two curves matches almost perfectly. Their maximum differences is at most $3.5$  units, and their maximum relative helicity difference is of 
$0.3\%$. The gauge invariance is thus very well respected for our estimation of the relative magnetic helicity.

In \fig{GaugecompEvolH}, right panel, the decomposition of the relative magnetic helicity is plotted. Unlike with the gauges of the 
practical DeVore method (cf. \fig{EvolEH}), $\Hp$ is now almost constantly null. It is not strictly null since we are not imposing 
$\vAp.\dS|_\surf =0$ on the side boundaries. We see that when the jet is developing $\Hp$ fluctuates slightly 
(barely visible on \fig{GaugecompEvolH}).  The helicity is thus distributed differently between $\Hm$, $\Hmix$ and $\Hp$ in the 
DeVore-Coulomb case compared to the practical DeVore gauge. As expected these helicity terms are indeed not gauge invariant. Most of the 
helicity that was carried by $\Hp$ for the practical DeVore gauge is now carried by $\Hm$ while $\Hmix$ remains very similar in both gauges 
(compare the green curves in \figs{EvolEH}{GaugecompEvolH}). In general, we have to expect a completely different distribution of 
$\Hm$, $\Hmix$ and $\Hp$ in other gauges. On the other hand, the quantity $\Hj$ and $\Hpj$ remains equal in both gauge computation 
(with the same precision as $H_\vol$). This is expected since $\Hj$ and $\Hpj$ are gauge-invariant quantities.

\begin{figure*}[ht]
  \setlength{\imsize}{0.35\textwidth}
  \sidecaption    
\includegraphics[width=\imsize,clip=true]{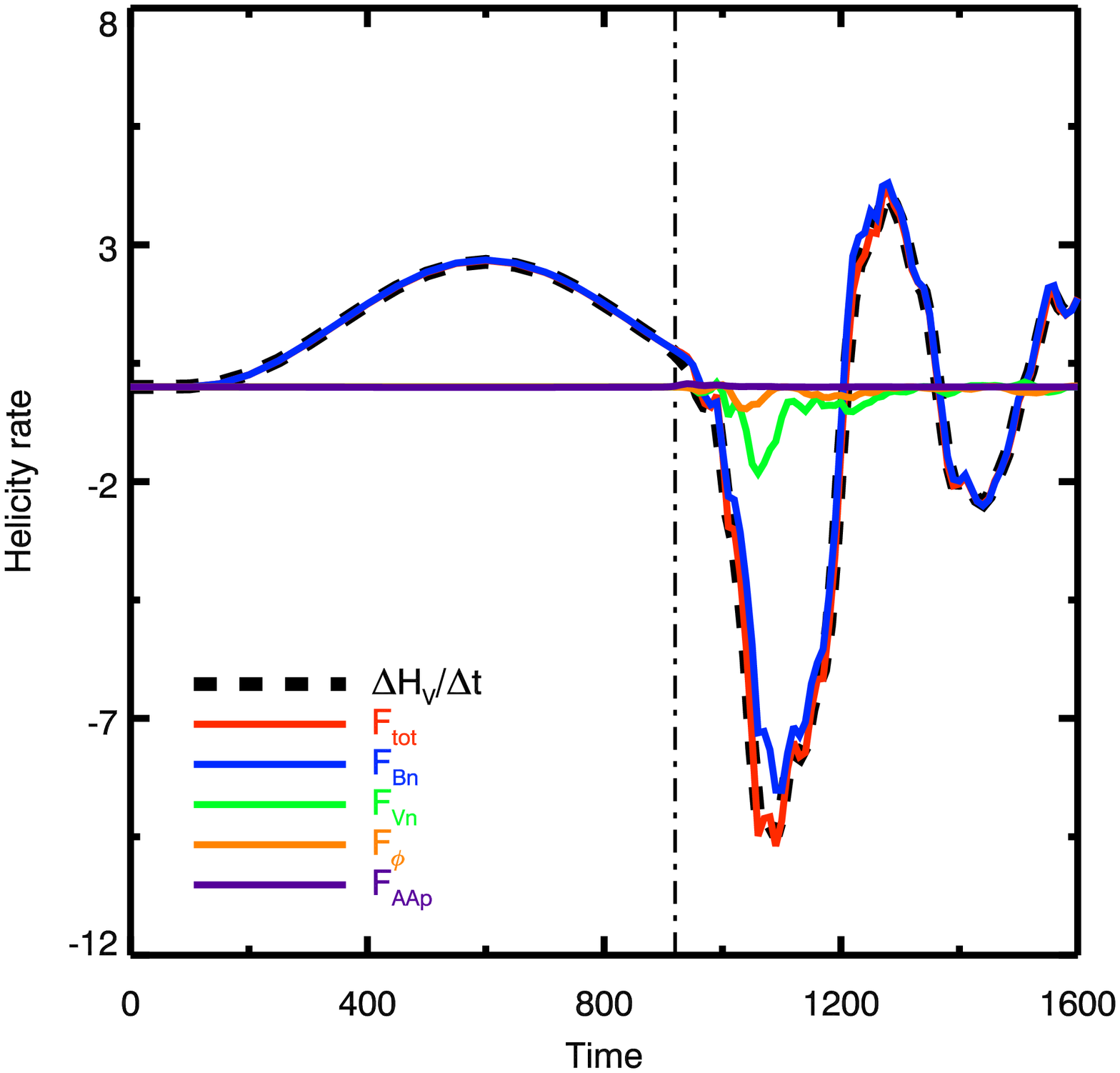}
 \includegraphics[width=\imsize,clip=true]{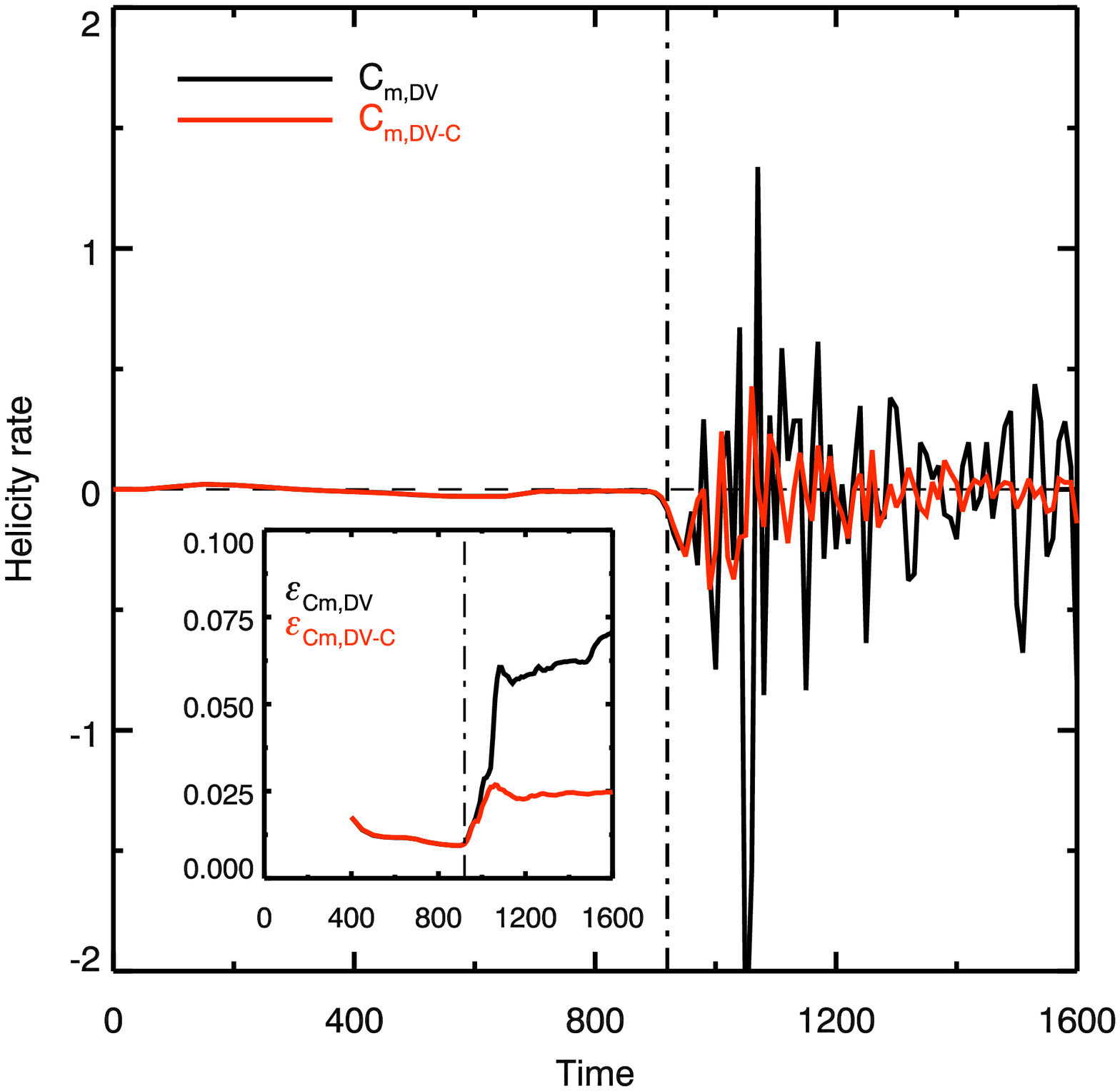}
  \caption{\textbf{Left panel}: Comparison of the helicity variation rate and the helicity flux integrated through the six boundaries of the domain 
  in the DeVore-Coulomb case. The plotted curves are similar than in Figure \ref{Fig:EvolHflux}.
  \textbf{Right panel}: magnetic helicity dissipation criterion $\Cm$, \eq{defCm}, plotted for the practical DeVore (black line) 
  and the DeVore-Coulomb (red line) methods. The insert shows 
$\epscm$, \eq{defepscm}, with the same color convention.
  }
  \label{Fig:GaugecompEvolHflux}
\end{figure*}

For the helicity fluxes, in the DeVore-Coulomb case the time variations $\Delta H_\vol/\Delta t$ of the helicity follows tightly the 
helicity flux through the boundary (\fig{GaugecompEvolHflux}, left panel). While $\Delta H_\vol/\Delta t$ is gauge invariant, the different 
contribution of $\Ftot$ are not. By comparison with \fig{EvolHflux} one observes that $\FBn$, $\FVn$, $\Fphi$ have a very different 
evolution. While $\Fphi$ was presenting fluctuations of the same amplitude as $\Ftot$ in the practical DeVore gauge, this term is now very weak 
throughout the evolution of the system with the DeVore-Coulomb method. The $\FBn$ term is dominating $\Ftot$ in the DeVore-Coulomb case 
both during the ideal MHD phase (at the bottom boundary) and during the non-ideal period (at the top boundary). $\FVn$ is weaker than with 
the practical DeVore gauge. The $\FAAp$ term remains negligible in both cases, although with a different choice of gauge, $\FAAp$ can contributes 
significantly to $\Ftot$ (see Appendix \ref{s:GenDev-Coul}).

The impact of the gauge dependance is more strikingly illustrated when looking at 
the time integrated helicity fluxes $H_{\surf,\#}$ (Eq.~\ref{eq:defHsurf}) through the boundaries. Their evolution is 
presented in \fig{GaugecompHVar} for the two gauges computations. While $H_\vol$ remains gauge invariant 
(\fig{GaugecompEvolH}, left panel), as expected the $H_{\surf,\#}$ present different profiles in the different gauges. $H_{\surf,\phi}$ is small 
when computed with the DeVore-Coulomb gauge while it presents large amplitudes in the practical DeVore gauge. $H_{\surf,Vn}$ and 
$H_{\surf,Bn}$ present smaller mean absolute values when $\vAp$ follows the DeVore-Coulomb conditions. 
Their ratio is also strongly dependent on the set of gauges employed. 

Numerous studies have computed $H_{\surf,Vn}$ \& $H_{\surf,Bn}$ and followed their time evolution in observed active 
regions \citep[e.g.][]{ZhangY12,LiuY13,LiuY14a,LiuY14b}. 
These terms have been incorrectly called "emergence" and "shear" terms and physical insight have 
been tried to be extracted from their values and respective ratio. However, as these terms are not gauge invariant, one must question the 
pertinence of such results. $H_{\surf,Vn}$ and $H_{\surf,Bn}$ cannot be studied independently as, for a given $\vv$, their intensities 
and respective values can be simply modified by a change of 
gauge.  This extends the conclusion of \citet{DemoulinBerger03} which showed that only the sum of 
$H_{\surf,Vn}$ and $H_{\surf,Bn}$ can be derived when only the tangential velocity components are known. 
More generally, the study presented here shows that, even when the full velocity 
field on the boundary is known, $H_{\surf,Vn}$ and $H_{\surf,Bn}$ can present different values depending on the gauge used for the computation. 
In summary, only the sum $H_{\surf}$ of all the flux terms, $H_{\surf,\#}$, is carrying a meaningful information.

\subsection{Helicity dissipation}\label{s:Dev-CoulHconservation}

The estimators $\Cr$ and $\Cm$ (cf. \sect{Mestimators}) enable us to study the helicity conservation and dissipation.
Since $\dHdt |_{\rm Bp,var}$ is null because of the Coulomb condition on the potential field,  $\Cm$ and $\Cr$
are equal to our numerical precision. There is no volume helicity variation due to the potential field. The conservation of the relative 
helicity is only limited by the actual dissipation helicity $\dHdt |_{\rm diss}$. The curve of $\Cr$  for the DeVore-Coulomb computation 
confirms that magnetic helicity is gauge-invariantly well conserved during the simulation (\fig{GaugecompEvolHflux}, right panel).  

In fact the computation in the DeVore-Coulomb case even improves our estimation of the helicity dissipation. 
During the quasi-ideal phase, the computation of $\Cm$ in the DeVore-Coulomb case is equal to its value with the practical DeVore case. 
However we observe that in the non-ideal phase $\Cm$ presents smaller oscillations when computed in the DeVore-Coulomb case. 
Theoretically $\Cm$, being equal to the helicity dissipation, $\dHdt |_{\rm diss}$, should be gauge invariant. However we observe that $\Cm$ can 
change by a factor 2 when computed with the different methods. With the practical DeVore method, the fluctuations of $\Cm$ peak at 25\% of 
the maximum amplitude of helicity variation during the non-ideal phase, while the peak is only equal to $4.5\%$ of the amplitude in the 
DeVore-Coulomb case. The non-dimensional criterion, $\epscm$ at the end of the simulation is equal to $7\%$ in the practical 
DeVore case while it is limited at $2.7\%$ with the DeVore-Coulomb method (see the inset in \fig{GaugecompEvolHflux}, right panel).

While the measure of $H_\vol$ is done with a high precision (see above), the estimation of the helicity flux, $\Ftot$ induces more numerical errors. 
The high frequency oscillations observed in the different terms contributing to $\Ftot$ are the symptom of the lower level of precision. 
The criterion $\Cm$ is in fact equal to the helicity dissipation plus the numerical errors in the volume helicity variations and in the helicity flux. 
What $\Cm$ is in fact providing is an upper value for the helicity dissipation. In the DeVore-Coulomb case, $\Cm$ is likely providing a smaller 
value, hence a better constraint on $\dHdt |_{\rm diss}$,  because the $\Ftot$ is dominated by the errors on $\FBn$, while in the practical 
DeVore case, the errors of the fluctuating $\FBn$, $\FVn$ and $\Fphi$ will sum up with comparable magnitudes.  

\begin{figure*}[ht]
  \setlength{\imsize}{0.35\textwidth}
  \sidecaption    
  \includegraphics[width=\imsize,clip=true]{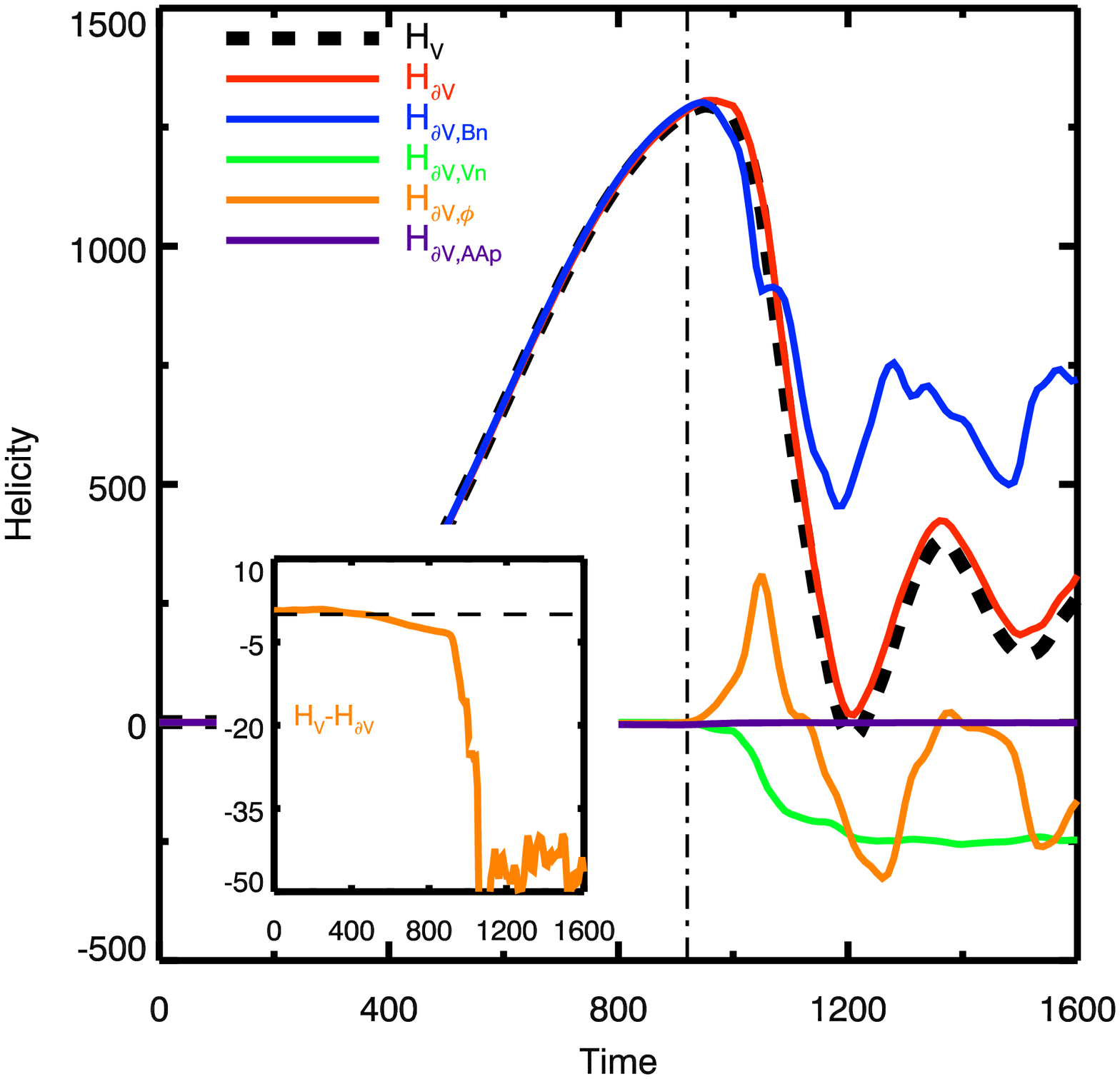}
  \includegraphics[width=\imsize,clip=true]{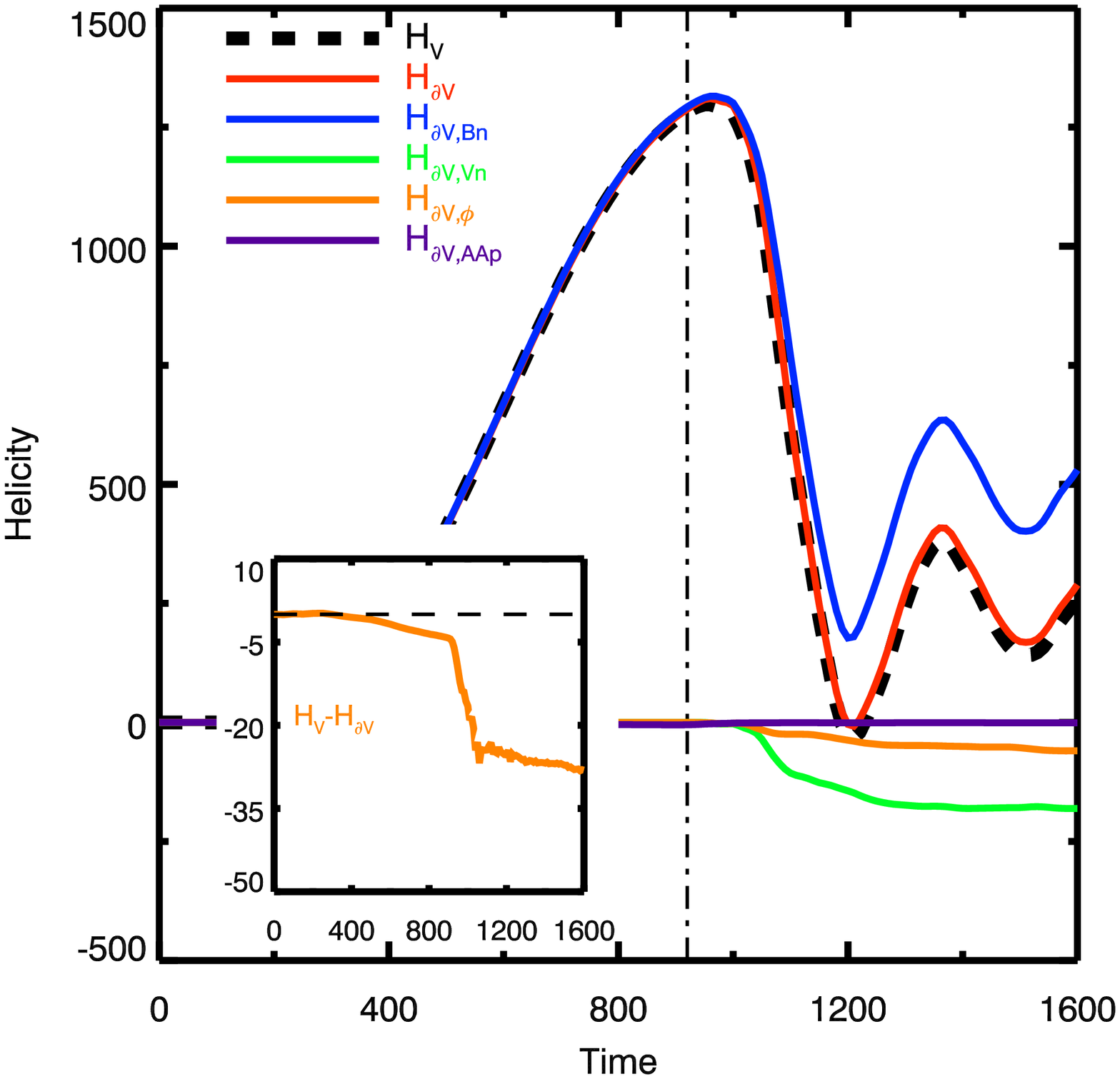}
  \caption{$H_\vol$ and $H_{\surf,\#}$ evolution in the system computed with different gauge. The left panel has been computed with the 
  practical DeVore method while the right panel has been derived with the DeVore-Coulomb method. The helicity variation 
  ($H_{\vol}$, black dashed line) is derived from the volume integration method. The time integrated helicity is $H_{\surf}$ (red line) which 
  can be decomposed in the contribution of its different terms:  $H_{\surf, \rm Bn}$ (blue line), $H_{\surf, \rm Vn}$ (green line), 
  $H_{\surf, \phi}$ (orange line) and $H_{\surf, \rm AAp}$ (purple line).  The orange line in the inserts correspond to the difference 
  between $H_\vol$ and $H_{\surf}$.
}
\label{Fig:GaugecompHVar}
\end{figure*}

The measure of the helicity flux through the boundary is the limiting factor that does not permit us to reach numerical precision during the 
non-ideal phase. It is likely that a higher spatial resolution sampling of the velocity field on the boundary sides will further improve 
the helicity dissipation estimation. Our present computation of $\Cm$ is therefore only an upper bound on the real helicity 
dissipation $\dHdt |_{\rm diss}$. Because of its lower $\epscm$, the DeVore-Coulomb method allows us to better bound the helicity dissipation. 

In DeVore-Coulomb computation, during the ideal-MHD phase, the maximum difference between $H_\vol$ and $H_\surf$ is of now of 
$4.3$ units and $\epsH (t=920) =0.3\%$. 
As theoretically expected, the dissipation of magnetic helicity is extremely small when only ideal process are present. It actually demonstrates that 
the ideality of the system is very well maintained by the numerical scheme during that phase. At the end of the non-ideal phase, 
the maximum difference between $H_\vol$ and $H_\surf$ is equal to $28.2$ units (orange line, \fig{GaugecompHVar}, right panel) and the 
relative amount of helicity dissipated is smaller than $\epsH(t=1600) =2.2\%$. 

In absolute value, the dissipation of magnetic helicity is 
thus very small compare to the magnetic helicity which is ejected away. The dissipated helicity represents less than $3\%$ of the helicity 
carried away by the jet. If one want to track the helicity evolution of this jet system, the helicity dissipation only represent a very minor 
contribution, more than one order of magnitude smaller than the helicity staying in the system and the ejected helicity. 
  
This low helicity dissipation is also to be compared to the relative decrease of magnetic energy of $62\%$ which is developing simultaneously. 
The reconnections generating the jets are thus dissipating/transforming magnetic energy $\gtrsim 30$ times more efficiently than magnetic helicity 
is dissipated. While vast amounts of energy are lost, magnetic helicity is barely affected by the non-ideal MHD processes at play.

\section{Conclusion} \label{s:Conclusion}

Based on the property that magnetic helicity presents an inverse cascade from small to large scale, \citet{Taylor74} conjectured that magnetic 
helicity, similarly to pure ideal MHD, is also effectively conserved when non-ideal processes are present. Because of the inherent difficulties to 
measure magnetic helicity, tests of this conjecture have so far been very limited (cf. \sect{intro}). The theoretical development of relative 
magnetic helicity \citep{BergerField84}, as well as the publication of recent methods to actually measure relative magnetic helicity in 
general 3D datasets \citep{Thalmann11,Valori12,YangS13} are now opening direct ways to test the conservation of magnetic helicity.

In this manuscript, we have performed the first precise and thorough test of \citet{Taylor74} hypothesis in a numerical simulation of 
an active solar-like event: the impulsive generation of a blowout jet \citep{Pariat09a}. Following \citet{YangS13}, our methods to 
test the level of magnetic helicity dissipation relies on the comparison of the variation of relative helicity in the domain with the fluxes of helicity 
through the boundaries.

This lead us to revise the formulation of the time variation of relative magnetic helicity in a fully bounded volume (cf. \sect{theoryHrelvar}). 
As relative magnetic helicity relies on 
magnetic vector potential, the question of the gauge is a central problematic of any magnetic-helicity related quantity. A general decomposition 
of the gauge-invariant time variation of magnetic helicity is given in \eq{Hvar4}, with no assumption made on the gauges of 
the magnetic field and of the reference potential field. Furthermore, we discussed how specific gauges and combination of gauges can simplify the 
formulation of the helicity variation (\sect{theoryHrelvarSpec}). 

Following \citep{Valori12}, we computed the variation of relative magnetic helicity using different gauges, 
all of them based on the gauge of \eq{DeVoreGauge} suggested by \citet{DeVore00}. We have been able 
to test the gauge dependance of several terms 
entering in the decomposition of relative magnetic helicity and its time variation. We demonstrated that all the quantities that were 
theoretically gauge-invariant ($H$, $\Hj$, $\Hpj$, $\dHdt$) were indeed invariant with a very good numerical precision ($<0.3\%$ of relative error).

Additionally, our analysis showed the effect of using different gauges on gauge-dependant quantities. Of particular interest are the results that we 
obtain on the $\FVn$ and $\FBn$ terms, \eqs{defFVn}{defFBn}, entering in the decomposition of the helicity flux. In some studies 
\citep[e.g.][]{LiuY14a,LiuY14b}, these terms have been used to putatively track the helicity contribution of vertical and horizontal plasma flows.  
However, our computations illustrates that these fluxes (and their ratio) vary with the gauges used hence precluding any meaningful physical 
insight of their interpretation in term of helicity. Following \citet{DemoulinBerger03}, we have concluded that only the total helicity flux, $\Ftot$, 
conveys a physical meaning.

Unlike magnetic helicity, even in ideal MHD, we showed that relative magnetic helicity cannot be expressed in general 
in classical conserved form. Only when computing the reference potential field with the Coulomb gauge, can the variation of the 
relative magnetic helicity be expressed in ideal MHD as a pure surface flux. Generally, relative magnetic helicity is therefore not 
a conserved quantity in a classical sense. 

In the first phase of our simulation, because of a topological constraint \citep{Pariat09a}, the system is believed to 
follow tightly an ideal MHD evolution. De facto, in our practical numerical case, we observed during that phase, 
that relative magnetic helicity is very well conserved, its variation following the time-accumulated flux of 
helicity with a relative accuracy of $0.3\%$. The relative helicity dissipation that we obtain is one order of magnitude smaller than the 
one estimated during the ideal evolution of  the simulation tested in \citet{YangS13}. The measure of the helicity dissipation can 
appear as a practical way to test the level of ideality in a simulation. During the ideal MHD phase, as expected, the 
Flux Corrected Transport scheme \citep{DeVore91} that is used to produce our test numerical simulation, 
is effectively able to ensure a quasi-ideal evolution with a measurable high degree of precision. 
 
Furthermore, the term-by-term study of the helicity variation enables to determine the real, gauge-invariant, dissipation of the 
magnetic helicity of the studied magnetic field $\dHdt_{\rm diss}$, Eq. (\ref{eq:defdHdiss}). 
For a solar-like active event, we have confirmed \citet{Taylor74} hypothesis that magnetic helicity is very well conserved even when 
non-ideal processes are acting. For the specific event that we have studied, even when intense 
magnetic reconnection is present, less than $2.2\%$ of the injected helicity is dissipated. While this is one order of magnitude larger 
than during the ideal phase, the dissipation of magnetic helicity is more than $30$ times smaller than the dissipation of magnetic energy 
during the same period.

\citet{YangS13} and this study are paving the way for future, more complete and more extensive tests of the Taylor's conjecture. 
In parallel to the exploration of the properties of helicity, our study offers also more numerically-oriented applications. 
In a previous work \citep{Valori13} we introduced a diagnostic for numerical discretization of magnetic field dataset that present finite error of 
non-solenoidality which are impacting the estimation of their magnetic energy. Further studies on the effect of time and spatial resolution, on a wider 
range of processes and dynamical MHD evolution are now needed. If precise 
quantitative bounds are placed on the level of helicity dissipation, magnetic helicity will eventually become useful to
test numerical MHD codes: the level of helicity dissipation could be used as a quantitative criterion of the quality of 
numerical experiments over the entire simulated evolution. With respect to the instantaneous divergence metric of the magnetic field, 
helicity is a complementary, extremely sensitive proxy suitable for testing integral conservation properties. 
The method of the measure of the helicity dissipation presented in this manuscript, would open up a new way to benchmark numerical codes.

The strong physical insight that can be gained by studying magnetic helicity is also further raised by our present study. More that forty years after, 
our numerically precise tests of \citet{Taylor74} conjecture on a solar like events, confirms that magnetic helicity is a quasi-conserved 
quantity of MHD. The application of the conservation of magnetic helicity is full of potential for the study of complex natural and 
experimental magnetised plasma systems. Because of its conservation, magnetic helicity may be the {\em raison d'\^etre} of the existence of 
coronal mass ejection \citep{Rust94,Low96}. Magnetic helicity can be tracked to characterise and relate the evolution of coronal active region with
interplanetary magnetic clouds \citep[e.g.][]{Mandrini05,Nakwacki11}. 

Finally, the impact of magnetic helicity conservation on the magnetic reconnection mechanism remains to be better understood \citep{Russell15}. 
While it has been observed that magnetic helicity can significantly modify the reconnection dynamics \citep{Linton01,DelSordo10}, 
how magnetic helicity is actually redistributed in the system, at quasi-constant total value by magnetic reconnection, still needs to be determined.

\begin{acknowledgements}
The authors dedicate this article to the French satirical magazine "Charlie Hebdo". This article was being redacted only few kilometres away 
and while "Charlie Hebdo" employees where assassinated because of their ideas. 
Freedom of though and of speech is the pedestal without which no true scientific research can be build.
GV thanks the Scientific Council of the Observatoire de Paris for supporting his stay during which this work has been initiated, and acknowledges the 
support of the Leverhulme Trust Research Project Grant 2014-051, and funding from the European Com missions Seventh Framework 
Programme under the grant agreements no. 284461 (eHEROES project). KD acknowledges funding from the Computational and Information 
Systems Laboratory, the High Altitude Observatory, and support from the Air Force Office of Scientific Research under award FA9550-15-1-0030. 
The National Center for Atmospheric Research is sponsored by the National Science Foundation.GV and EP thanks ISSI, its members and 
the participants of the ISSI international team on {\it Magnetic 
Helicity estimations in models and observations of the solar magnetic field} where this work has been discussed and commented. 
The numerical simulations presented in this article have been performed 
thanks to the HPC resources of CINES, granted under the allocations 2014--046331 by GENCI (Grand Equipement National de Calcul Intensif).   
\end{acknowledgements}

 \bibliographystyle{aa}  
\bibliography{HCons}       
\IfFileExists{\jobname.bbl}{}  
{ 
\typeout{} 
\typeout{****************************************************} 
\typeout{****************************************************} 
\typeout{** Please run "bibtex \jobname" to obtain}  
\typeout{**the bibliography and then re-run LaTeX}  
\typeout{** twice to fix the references!} 
\typeout{****************************************************} 
\typeout{****************************************************} 
\typeout{} 
 }

\begin{appendix}

\section{Helicity test without the condition: $\vA (z_{\rm top}) = \vAp(z_{\rm top})$} \label{s:GenDev-Coul}

\begin{figure*}[ht]
  \setlength{\imsize}{0.35\textwidth}
  \sidecaption    
  \includegraphics[width=\imsize,clip=true]{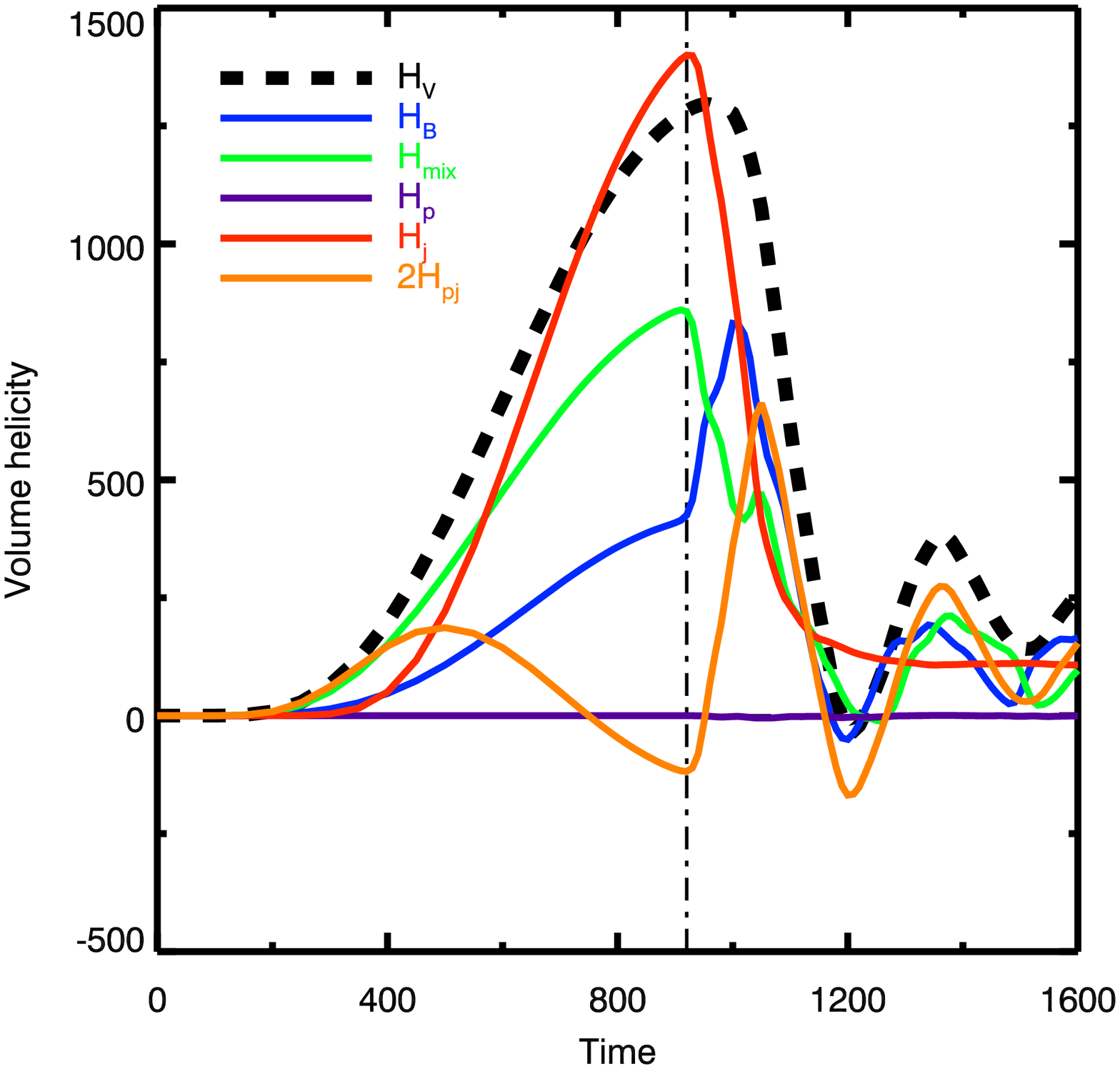}
  \includegraphics[width=\imsize,clip=true]{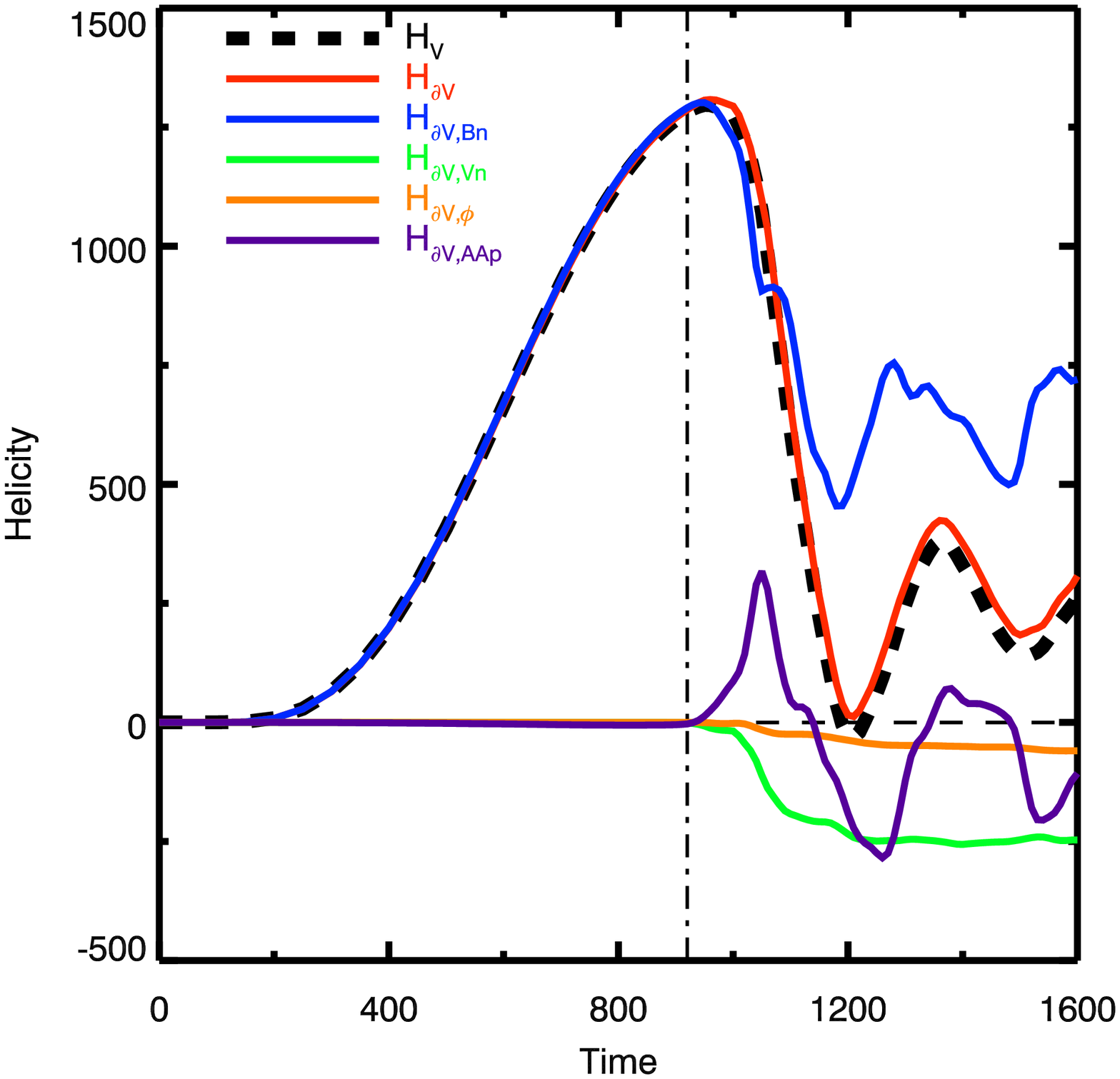}
  \caption{ \textbf{Left panel}: Relative magnetic helicity ($H_{\vol}$, black dashed line) and its decomposition in the general DeVore-Coulomb method. The plotted quantities are the same as in  Figures \ref{Fig:EvolEH}, bottom and \ref{Fig:GaugecompEvolH}, right panels.
   \textbf{Right panel}: $H_\vol$ and $H_{\surf,\#}$ evolution in the system computed with the general DeVore-Coulomb method. The plotted quantities are the same as in \fig{GaugecompHVar}. 
  }
\label{Fig:GenDV-C}
\end{figure*}

In the numerical implementation of both the practical DeVore method and the DeVore-Coulomb methods, we impose the condition 
of \eq{CondAAptop}, of having the same distribution of $\vA$ and $\vAp$ at the top boundary, in addition of the DeVore gauge 
(Eq.~\ref{eq:DeVoreGauge}) which is valid in the whole volume. Because of this condition, $\vA$ and $\vAp$ are both different 
when computed with one method or the other. The gauge of $\vA$ and the gauge of $\vAp$ are linked by  \eq{CondAAptop}. It
induces that $\FAAp$ is null at the top boundary, and de facto reduces its intensity in the whole domaine

It is however possible to compute and estimate the helicity without enforcing \eq{CondAAptop}. We can follow the helicity evolution by mixing 
the vector potential computed with both method. Since gauge invariance of \eq{defHrel} 
does not require the use of the same gauge for $\vA$ and $\vAp$,
we can use the $\vA$ computed with the practical DeVore method with $\vAp$ derived 
with the DeVore-Coulomb methods. $\vAp$ is thus still satisfying both the DeVore and the Coulomb gauge. 
As $\vA$, which is only satisfying the DeVore gauge condition, has been computed independently of $\vAp$, there is no boundary 
surface along which they share any common distribution. In this appendix we refer to this derivation as the "general DeVore-Coulomb" case.

\fig{GenDV-C}, right panel, presents the different terms entering in the decompositions \eqs{HrelDecomp}{HrelDecomp2} of 
the relative helicity. As in \sect{Dev-CoulHevolution}, the gauge invariance of $H_\vol$ in this computation relatively to the other 
methods is ensured with a high precision (<$0.3\%$). As with the others methods, one remarks that $\Hj$ and $\Hpj$ remains constant while 
$\Hm$, $\Hmix$ and $\Hp$ are different in the general DeVore-Coulomb case. This further confirms the gauge dependance properties of 
each decomposition.

When looking at the time integrated helicity fluxes (\fig{GenDV-C}, right panel), we find again that $H_{\surf,tot}$ is 
following tightly the variation of helicity $H_\vol$. As for the two other methods, the helicity dissipation is also very small and with a precision 
similar to the practical DeVore case (\sect{DevHconservation}). What significantly differs from the two other methods is 
the repartition of the helicity flux $\Ftot$ between the different terms which composes it.

Since $\vAp$ is respecting the Coulomb condition the term $\dHdt_{\rm p,var}$ is null to the numerical precision. Since $\Fphi$ only 
involves quantities based on the derivation of the potential field, $H_{\surf,\phi}$ are equal for both the DeVore-Coulomb and the 
general DeVore-Coulomb cases. On the other hand, as $\FBn$, $\FVn$ are only involving $\vA$,  
$H_{\surf,Vn}$ and $H_{\surf,Bn}$ in the general DeVore-Coulomb are equal with their respective estimations in the practical DeVore case.

It is therefore $\FAAp$ with concentrate the helicity flux contribution that enable $\Ftot$ to be quasi gauge-invariant for the three derivations 
(\fig{GenDV-C}, right panel). While 
$H_{\surf,AAp}$ was negligeable in both the practical DeVore and the DeVore-Coulomb cases, we observe that this term is now a major 
contributor of the helicity fluxes. This is not surprising since $H_{\surf,AAp}$ results from the existence of large differences between the 
distribution of $\vA$ and $\vAp$ on the boundaries. Both the computations in the practical DeVore and the DeVore-Coulomb methods 
were enforcing \eq{CondAAptop}, which induces a very weak value of $H_{\surf,AAp}$. We observe that dropping the condition 
(\ref{eq:CondAAptop}) creates a strong $H_{\surf,AAp}$.

This test further demonstrates that the choice of the gauge strongly influences the distribution of the helicity fluxes
composing the total helicity flux $\Ftot$. Only the total flux $\Ftot$ is a quasi gauge-invariant. None of the terms which are composing the 
helicity flux $\Ftot$ shall a priori be neglected. Depending on the gauge, each term can carry a significant contribution. In a numerical estimation, it 
is thus highly advisable to compute all the terms which are forming the helicity flux density (Eq.~\ref{eq:Hvar4}). Explicitly computing each terms 
allows us to verify that the constraints set on the used gauges are effectively enforced numerically.

\end{appendix}

 \end{document}